\documentclass[a4paper,oneside,11pt]{article}

\usepackage[margin=26mm]{geometry}
\usepackage[utf8]{inputenc}
\usepackage[caption=false]{subfig}
\usepackage{enumitem}
\usepackage{colortbl}
\usepackage{tabularx}
\usepackage{listings}
\usepackage{algpseudocode}
\usepackage{algorithm}
\usepackage{xcolor}
\usepackage{amsmath}
\usepackage{graphicx}
\usepackage{amssymb}
\usepackage{authblk}
\usepackage[
backend=bibtex,
safeinputenc,
bibencoding=utf8,
sorting=none,
maxcitenames=10,
maxbibnames=10,
style=numeric,
citestyle=numeric-comp]{biblatex}
\usepackage{lastpage}

\usepackage{fancyhdr}
\begin{document}

	\author[1]{Sebastian Ruland}
	\author[2]{Malte Lochau}
	\affil[1]{Technical University Darmstadt, Darmstadt, Germany}
	\affil[2]{University of Siegen, Siegen, Germany}
	
	\title{On the Interaction between Test-Suite Reduction and Regression-Test Selection Strategies}
	

	
	\maketitle
	\begin{abstract}
		Unit testing is one of the most established quality-assurance techniques for software development.
		One major advantage of unit testing is the adjustable
		trade-off between efficiency (i.e., testing effort) and 
		effectiveness (i.e., fault-detection probability).
		To this end, various strategies have been proposed to exploit this trade-off.
		In particular, test-suite reduction (TSR) reduces the number of (presumably redundant) test cases while testing a single program version.
		Regression-test selection (RTS) selects test cases for testing consecutive program revisions.
		However, both TSR and RTS
		may influence---or even obstruct---
		each others' performance when used in combination.
		For instance, test cases discarded during TSR for a particular program version 
		may become relevant again for RTS.
		However, finding a combination of both strategies leading to a reasonable trade-off throughout the version history of
		a program is an open question.
		The goal of this paper is to gain a better understanding of the interactions between TSR and RTS with respect
		to efficiency and effectiveness.
		To this end, we present a configurable framework called \textsc{RegreTS}
		for automated unit-testing of C programs. The framework comprises different
		strategies for TSR and RTS
		and possible combinations thereof.
		We apply this framework to a collection of subject systems,
		delivering several crucial insights. First, TSR has almost always a negative
		impact on the effectiveness of RTS, yet
		a positive impact on efficiency. Second, test cases revealing to testers the effect of program modifications between consecutive
		program versions are far more effective than test cases simply
		covering modified code parts, yet causing much more testing
		effort.
	\end{abstract}


	\thispagestyle{empty}

	\newsavebox{\cprogram}
	\newsavebox{\cprogramPatchOne}
	\newsavebox{\cprogramPatchTwo}
	\newsavebox{\cprogramPatchThree}
	\newsavebox{\patchOne}
	\newsavebox{\patchTwo}
	\newsavebox{\patchThree}
	\newsavebox{\compareMR}
	\newsavebox{\compareMT}
	\newsavebox{\compareInvalid}
	\begin{lrbox}{\cprogram}
		\centering
		\begin{minipage}{\textwidth}
			\begin{lstlisting}
				int find_last (int x[], int y) {'\label{line:function-start_p0}'
					if(x[0] <= 0)'\label{line:if-statement_p0}'
					return -1;'\label{line:return-1_p0}'
					int last = -2;
					for (int i=0; i <= x[0]-2; i++) '\label{line:for-bug_p0}'
					if (x[i] <= y)'\label{line:if-bug_p0}'
					last = i;'\label{line:lastassign_p0}'
					return last;'\label{line:return-last_p0}'
				}
			\end{lstlisting}
		\end{minipage}
	\end{lrbox}
	\begin{lrbox}{\cprogramPatchOne}
		\centering
		\begin{minipage}{\textwidth}
			\begin{lstlisting}
				int find_last (int x[], int y) {'\label{line:function-start_p1}'
					if(x[0] <= 0)
					return -1;'\label{line:return-1_p1}'
					int last = -2;
					for (int i=1; i <= x[0]-2; i++) '\label{line:for-bug_p1}'
					if (x[i] <= y)'\label{line:if-bug_p1}'
					last = i;'\label{line:return-i_p1}'
					return last;
				}
			\end{lstlisting}
		\end{minipage}
	\end{lrbox}
	\begin{lrbox}{\cprogramPatchTwo}
		\begin{minipage}{\textwidth}
			\begin{lstlisting}
				int find_last (int x[], int y) {'\label{line:function-start_p2}'
					if(x[0] <= 0)
					return -1;'\label{line:return-1_p2}'
					int last = -2;
					for (int i=1; i <= x[0]-1; i++) '\label{line:for-bug_p2}'
					if (x[i] <= y)'\label{line:if-bug_p2}'
					last = i;'\label{line:return-i_p2}'
					return last;
				}
			\end{lstlisting}
		\end{minipage}
	\end{lrbox}
	
	\begin{lrbox}{\cprogramPatchThree}
		\begin{minipage}{\textwidth}
			\begin{lstlisting}
				int find_last (int x[], int y) {'\label{line:function-start_p3}'
					if(x[0] <= 0)
					return -1;'\label{line:return-1_p3}'
					int last = -2;
					for (int i=1; i <= x[0]-1; i++) '\label{line:for-bug_p3}'
					if (x[i] == y)'\label{line:if-bug_p3}'
					last = i;'\label{line:return-i_p3}'
					return last;
				}
			\end{lstlisting}
		\end{minipage}
	\end{lrbox}
	\begin{lrbox}{\patchOne}
		\begin{minipage}{\textwidth}
			\begin{lstlisting}[numbers=none]
				'\ref{line:for-bug_p0}-\phantom{}-: 'for (int i=0; i <= x[0]-2; i++
				'\ref{line:for-bug_p1}++: 'for (int i=1; i <= x[0]-2; i++
			\end{lstlisting}
		\end{minipage}
	\end{lrbox}
	\begin{lrbox}{\patchTwo}
		\begin{minipage}{\textwidth}
			\begin{lstlisting}[numbers=none]
				'\ref{line:for-bug_p1}-\phantom{}-: 'for (int i=1; i <= x[0]-2; i++
				'\ref{line:for-bug_p2}++: 'for (int i=1; i <= x[0]-1; i++
			\end{lstlisting}
		\end{minipage}
	\end{lrbox}
	
	\begin{lrbox}{\patchThree}
		\begin{minipage}{\textwidth}
			\begin{lstlisting}[numbers=none]
				'\ref{line:if-bug_p2}-\phantom{}-: 'if (x[i] <= y)
				'\ref{line:if-bug_p3}++: 'if (x[i] == y)
			\end{lstlisting}
		\end{minipage}
	\end{lrbox}
	
	\begin{lrbox}{\compareMR}
		\begin{minipage}{\textwidth}
			\begin{lstlisting}[numbers=none]
				void find_last_p2_3(int x[], int y){
					if(find_last_p2(x,y) != find_last_p3(x,y)
					test_goal:printf("differencing test case found");'\label{line:tg_revealing}'
				}
			\end{lstlisting}
		\end{minipage}
	\end{lrbox}
	
	\begin{lrbox}{\compareMT}
		\begin{minipage}{\textwidth}
			\begin{lstlisting}
				int find_last_p2_3(int x[], int y) {
					if(x[0] <= 0)
					return -1;'\label{line:return-1_p_comp}'
					
					int last = -2;
					for (int i=1; i <= x[0]-1; i++) '\label{line:for-bug_p_comp}'{
						test_goal:'\label{line:tg_comp}'
						if (x[i] == y)'\label{line:if-bug_p_comp}'
						last = i;'\label{line:return-i_p_comp}'
					}
					return last;
				}
			\end{lstlisting}
		\end{minipage}
	\end{lrbox}
	
	\begin{lrbox}{\compareInvalid}
		\begin{minipage}{\textwidth}
			\begin{lstlisting}
				int find_last_p3 (int x[], int y) {...}
				struct intStruct find_last_p4 (int x[], int y) {...}
				
				
				void find_last_p3_4(int x[], int y){
					if(find_last_p3(x,y) != find_last_p4(x,y)
					test_goal:printf("differencing test case found");
				}
			\end{lstlisting}
		\end{minipage}
	\end{lrbox}
	
	\section{Introduction}\label{sec1:intro}
	\paragraph{Background and Motivation.}
	Software testing is concerned with revealing
	as many bugs as possible 
	in a program within a---usually strictly limited---amount of time~\cite{Ammann2016}.
	In particular, \emph{unit testing} is one of the most important 
	innovations in the recent past for pro-actively ensuring software quality
	in an effective, yet tractable and agile manner.
	Bugs revealed by unit tests include program crashes caused by programming
	errors as well as faulty input-output value pairs contradicting a given specification 
	of the expected behavior (e.g., functionally incorrect program logics or ambiguous requirements).
	To this end, \emph{test cases} are defined in terms of exemplary input values and expected output values
	for experimentally executing the program unit under test in a systematic manner.

	One advantage of software testing as compared to other
	quality-assurance techniques is the---more or less---freely adjustable 
	trade-off between two (generally conflicting) optimization goals, namely
	\emph{effectiveness} (i.e., maximizing fault-detection probability) and \emph{efficiency}
	(minimizing testing effort)~\cite{Yoo2012,Engstrom2008,Harrold1993}.
	In an idealistic setting, a perfect trade-off between both goals would be 
	a test suite that finds \emph{all} bugs with \emph{minimum} effort (e.g., requiring a 
	minimum number of test cases).
	In reality, however, the number and exact location of bugs are, unfortunately,
	a-priori unknown; and even if all bugs would be known in advance (which would make testing obsolete), 
	then finding a minimum set of test cases for revealing them is NP-hard~\cite{Khalilian2012}.
	Various \emph{heuristics} have been proposed to control the selection of
	sufficiently effective, yet efficient sets of test cases for a program unit under test.
	Heuristics for measuring effectiveness 
	often rely on structural \emph{code-coverage metrics} (e.g.,
	branch coverage) to be optimized by a selected test suite~\cite{Anand2013}, whereas heuristics for efficiency 
	apply test-suite reduction (TSR) techniques to decrease the number of \emph{redundant} 
	test cases, again, with respect to that given code-coverage criterion~\cite{Yoo2012}.

	Moreover, modern software development is faced with the challenge of
	ever-shortened release cycles leading to
	increasing frequencies of consecutive program revisions. 
	The goal of \emph{regression testing} is to reveal emerging bugs introduced
	by erroneous program modifications between subsequent program versions~\cite{Yoo2012}.
	RTS strategies define criteria for
	updating an existing test suite of a previous program version,
	by removing outdated test cases, by keeping still relevant test cases, 
	and by adding further test cases for newly added functionality.
	Many RTS and regression-test generation strategies are
	further concerned with prioritizing test cases 
	to apply the presumably more effective test cases first.
	The aforementioned trade-off between 
	efficiency and effectiveness is therefore also the primary goal of 
	regression testing, but now starting from 
	an existing test suite inherited from previous program revisions.

	\paragraph{Problem Statement and Research Challenges.}
	Both \emph{TSR strategies} for testing a single program
	version as well as \emph{regression-testing strategies} for testing
	consecutive program revisions are concerned with the same problem: 
	\begin{quotation}
		\emph{How to achieve a reasonable trade-off between efficiency and effectiveness?}
	\end{quotation}
	
	Nevertheless, both kinds of strategies may potentially influence---or even obstruct---
	each other in various ways.
	For instance, a test case being considered redundant for one program
	version (thus being removed during TSR) may become
	relevant, again, for later program revisions.
	As a result, expensive re-selections of eventually ``lost'' test cases may become necessary
	during regression testing.
	Hence, excessive TSR, although potentially
	improving testing efficiency for one program version, may,
	eventually, have a negative impact on regression testing.
	On the other hand, too reluctant TSR
	may lead to an ever-growing test suite thus also increasing test-selection effort during regression testing.
	Those subtle interactions between TSR and regression testing
	are neither obvious nor fully predictable in advance.
	Finding a feasible combination of both strategies yielding a suitable
	efficiency/effectiveness trade-off throughout the entire version history of
	a program is an open challenge involving further fine-grained details to be taken into account
	(e.g., code-coverage criteria, TSR heuristics, RTS criteria etc.).

	\paragraph{Contributions.}
	We present a conceptual framework for an in-depth
	investigation of interactions between strategies for TSR and
	RTS and the resulting 
	impact on (mutually contradicting) efficiency and effectiveness measures.
	We focus on unit testing of C programs 
	and use practically established control-flow coverage criteria as 
	effectiveness measure for RTS and TSR~\cite{Harrold1993}.
	As effectiveness heuristics for RTS, we consider 
	modification-traversing as well as modification-revealing test cases~\cite{Yoo2012}.

	In our experimental comparison of the 
	different strategies, we additionally consider fault-detection rate
	as a-posteriori effectiveness measure.
	Concerning testing efficiency, we measure sizes of test suites (i.e., number of test cases)
	as well as the computational effort (i.e., CPU-time) for (regression) test-case generation and/or selection as well as
	TSR.

	To summarize, we make the following contributions.
	\begin{itemize}
		\item \textbf{Configurable framework} for unit-test generation and selection,
		integrating recent TSR strategies, as well as existing and novel RTS strategies.
		
		\item \textbf{Tool support} for automatically applying different strategies
		to a version history of C program units.
		
		\item \textbf{Experimental evaluation results} gained from applying our
		tool to a collection of C program units with available version history.
		The evaluation results show that:
		
		(1) TSR based on a coverage-based notion of test-case redundancy
		almost always decreases effectiveness of RTS, yet
		obviously having a positive impact on testing efficiency,
		
		(2) modification-revealing test cases are far more effective than modification-traversing 
		test cases for RTS, yet modification-revealing test-case generation requires much more computational effort and
		
		(3) the number of test cases and the number of previous program versions considered
		for RTS only has a low impact on the effectiveness of regression testing.	
	\end{itemize}
	
	\paragraph{Outline.}
	The remainder of this paper is structured as follows.

	\noindent\textbf{Section 2} contains an overview on the necessary background
	and terminology used in the remainder of the paper.
	We first introduce an illustrative example to introduce essential
	notions and concepts and also use this example to derive the motivation
	for the methodology proposed in the main part of the paper.
	
	\noindent\textbf{Section 3} contains a conceptual description of 
	the proposed evaluation framework for investigating the interaction
	between TSR and RTS strategies.
	After a general overview, the different possible strategies are then described in more detail
	using the illustrative example.
	
	\noindent\textbf{Section 4} provides a detailed description of our tool support
	for the envisioned methodology together with experimental
	evaluation results gained from applying the tool to a collection
	of subject systems.
	
	\noindent\textbf{Section 5} provides an overview about related work
	from the fields of regression testing strategies, automated test-case generation, 
	as well as related approaches for regression verification.
	
	\noindent\textbf{Section 6} concludes the paper and provides
	a brief outlook on possible future work.
	
	\paragraph{Verifiability.}
	To make our results reproducible, we provide the
	tool implementation and all experimental results as well as raw
	data on a supplementary web page\footnote{https://www.es.tu-darmstadt.de/regrets}.

	\section{Background and Motivation}\label{sec2:background}
	We first describe the necessary background for the rest of this paper.
	We introduce an illustrative running example by means of a 
	small, evolving C program unit with corresponding unit test cases.
	Based on this example, we describe basic notions of
	unit testing, test-suite reduction and regression testing.
	We conclude by summarizing the research
	challenges addressed in the remainder of this paper.

	\paragraph{Program Units.}
	Let us consider the sample C program
	unit in Fig.~\ref{fig:c-program}.
	This source code~\footnote{We will use the terms program, source code and program unit interchangeably for convenience} 
	constitutes the initial version $P_0$ of 
	an evolving program developed and (re-)tested in an incremental manner.
	Function \textbf{\texttt{find\_last}} receives as inputs an integer-array
	\textbf{\texttt{x[]}} and an integer value \textbf{\texttt{y}} and
	is supposed to return the index of the \emph{last occurrence} of the value of 
	\textbf{\texttt{y}} in \textbf{\texttt{x[]}}.
	The first element of the array is supposed to carry as meta-information the \emph{length} 
	(i.e., the number of elements) of the array. 
	For this reason, the first element should be ignored during the search for the value of \textbf{\texttt{y}}.
	The search itself is supposed to be performed by iterating over the array and by 
	keeping track of the latest occurrence of \textbf{\texttt{y}}~\footnote{Of course, a more reasonable implementation
		may perform a reversed iterative search and return the first occurrence. The purpose
		of the example is to demonstrate essential testing concepts in a condensed way thus requiring
		some simplifications which do, however, not threaten the validity of the overall approach.}.
	Additionally, the function should return predefined error codes:
	\begin{itemize}
		\item If the size of the array (i.e., the value of the first element) 
		is less or equal to zero, the returned error-code is \textbf{\texttt{-1}}.
		\item If the value of \textbf{\texttt{y}} does not occur
		in the array, the returned error-code is \textbf{\texttt{-2}}. 
	\end{itemize}
	\begin{figure}[tp]
		\centering
		\subfloat[Initial Version $P_0$ of a Program Unit]{\label{fig:c-program}\usebox{\cprogram}}
		
		\centering
		\subfloat[First Bug Fix]{\label{fig:c-program-fix1-patch}\parbox{\linewidth}{\centering\usebox{\patchOne}}}
		
		\centering
		\subfloat[New Program Version $P_1$ After Applying the First Bug Fix]{\label{fig:c-program-fix1-program}\parbox{\linewidth}{\centering\usebox{\cprogramPatchOne}}}
		
		\centering
		\subfloat[Second Bug Fix]{\label{fig:c-program-fix2-patch}\parbox{\linewidth}{\centering\usebox{\patchTwo}}}
		
		\centering
		\subfloat[New Program Version $P_2$ After Applying the Second Bug Fix]{\label{fig:c-program-fix2-program}\parbox{\linewidth}{\centering\usebox{\cprogramPatchTwo}}}
		
		\centering
		\subfloat[Third Bug Fix]{\label{fig:c-program-fix3-patch}\parbox{\linewidth}{\centering\usebox{\patchThree}}}
		
		\centering
		\subfloat[New Program Version $P_3$ After Applying the Third Bug Fix]{\label{fig:c-program-fix3-program}\parbox{\linewidth}{\centering\usebox{\cprogramPatchThree}}}
		
		\centering
		\caption{Program Versions $P_0$, $P_1$, $P_2$ and $P_3$ with their corresponding Bug Fixes}
		\label{fig:c-program-fixes}
	\end{figure}

	\paragraph{Program Bugs.}
	The initial program version $P_0$ in Fig.~\ref{fig:c-program} 
	does, however, not satisfy the specified functionality as it contains the following three \emph{bugs}.
	\begin{enumerate}
		\item \emph{Bug 1}: The search index starts at $0$, thus incorrectly 
		including the value of the meta-information into the search (see line~\ref{line:for-bug_p0}).
		\item \emph{Bug 2}: The search index stops at $x[0] - 2$ thus incorrectly 
		excluding the last element from the search (see line~\ref{line:for-bug_p0}).
		\item \emph{Bug 3}: The search incorrectly matches all values smaller than, or equal to, \textbf{\texttt{y}} 
		instead of solely considering values equal to \textbf{\texttt{y}} (see line~\ref{line:if-bug_p0}).
	\end{enumerate}
	To fix those bugs, assume the developer to consecutively
	creates \emph{program revisions} by correcting erroneous code parts.

	\paragraph{Program Revisions.}
	Let us assume that three consecutive \emph{revisions}, $P_1$, $P_2$  and $P_3$, of the initial program version $P_0$
	have been created, where $P_3$ finally has all three bugs fixed as described above.
	A \emph{program revision} denotes a \emph{new program version} as a result of \emph{modifications} 
	made to some parts of the previous program version.
	We represent program revisions using the established \textsc{diff}-syntax frequently used in patch files.
	Line numbers marked with post-fix \textbf{\texttt{-\phantom{}-}} refer to the respective lines in 
	the previous program version being removed in the new version, and those with
	post-fix \textbf{\texttt{++}} refer to lines in the new version 
	being added to the previous version:
	\begin{itemize}
		\item \emph{Patch 1} in Fig.~\ref{fig:c-program-fix1-patch} provides
		a fix for Bug 1.
		\item \emph{Patch 2} in Fig.~\ref{fig:c-program-fix2-patch} provides
		a fix for Bug 2.
		\item \emph{Patch 3} in Fig.~\ref{fig:c-program-fix3-patch} provides
		a fix for Bug 3.
	\end{itemize}
	After the first bug fix, we obtain the improved (yet still faulty) program version $P_{1}$ 
	(cf. Fig.~\ref{fig:c-program-fix1-program}). 
	The second bug fix to $P_{1}$ (yielding program version $P_{2}$, cf. Fig.~\ref{fig:c-program-fix2-program}) and the third bug fix to $P_{2}$ 
	finally yield the bug-free version $P_{3}$ (cf.~Fig.~\ref{fig:c-program-fix3-program}).
	Such an incremental process is often interleaved
	by consecutive \emph{unit-testing steps} to assure correctness of new versions and/or 
	to reveal further bugs potentially emerging during revisions.

	\paragraph{Unit Testing.}
	Test cases for unit testing by means of program inputs are usually selected with respect to (structural) 
	\emph{code coverage criteria}.
	For instance, we require at least two \emph{different} test cases for \emph{branch coverage} 
	of program version $P_0$.
	First, we require a test case for covering (reaching) the \textbf{\texttt{true}}-branch of the \textbf{\texttt{if}}-statement in line~\ref{line:if-statement_p0} (i.e., the input-array is empty).
	Additionally, we require at least one further test case for covering the remaining branches as follows.
	\begin{itemize}
		\item The \textbf{\texttt{false}}-branch of the \textbf{\texttt{if}}-statement in line~\ref{line:if-statement_p0} (i.e., requiring the input-array to have at least one element).
		\item The \textbf{\texttt{true}}-branch of the \textbf{\texttt{for}}-loop in line~\ref{line:for-bug_p0} 
		(i.e., requiring the input-array to have at least two elements).
		\item The \textbf{\texttt{false}}-branch of the \textbf{\texttt{for}}-loop in line~\ref{line:for-bug_p0} 
		(i.e., requiring the input-array to have at least one element).
		\item The \textbf{\texttt{true}}-branch of the \textbf{\texttt{if}}-statement in line~\ref{line:if-bug_p0} 
		(i.e., requiring the input-array to have at least one element being less or equal to \textbf{\texttt{y}}).
		\item The \textbf{\texttt{false}}-branch of the \textbf{\texttt{if}}-statement in line~\ref{line:if-bug_p0} 
		(i.e., requiring the input-array to have at least one element not being less or equal to \textbf{\texttt{y}}).
	\end{itemize}
	
	To satisfy branch coverage on $P_0$, a developer/tester may select a \emph{test suite} 
	consisting of the following two \emph{test cases}\footnote{More precisely, this test suite is only 
		able to reach Line 8 due to Bug 3}:
	\begin{itemize}
		\item $t_1$= (\textbf{\texttt{x=[0], y=0}}),
		\item $t_2$= (\textbf{\texttt{x=[3,5,5,3], y=4}}).
	\end{itemize}
	We denote test cases $t$ as collections of input-value assignments
	(i.e., an array \textbf{\texttt{x}} and a value for \textbf{\texttt{y}}). 
	Test-case specifications are often further equipped with the expected output
	values (i.e., $\textbf{\texttt{last=-1}}$ for $t_1$ and $\textbf{\texttt{last=-2}}$ for $t_2$).
	If applied to $P_0$, $t_1$ would \emph{pass}, whereas $t_2$ would indeed \emph{fail} 
	as it produces the erroneous return value \textbf{\texttt{0}} 
	instead of the expected value \textbf{\texttt{-2}}. 
	Hence, this test suite, although satisfying branch coverage, 
	only reveals Bug~1 and Bug~3, but not Bug~2.
	In contrast, a test suite containing as test cases:
	\begin{itemize}
		\item $t_1$= (\textbf{\texttt{x=[0], y=0}}),
		\item $t_3$= (\textbf{\texttt{x=[1,1,1], y=2}}),
		\item $t_4$= (\textbf{\texttt{x=[1,2,2], y=0}}).
	\end{itemize}
	also satisfies branch coverage and reveals Bug~1 and Bug~2, but not Bug~3.
	Hence, this test suite is similarly \emph{effective} as the first (revealing two out of three bugs), yet 
	being less \emph{efficient} as it causes more testing effort due to the additional test-case execution. 
	The test suite:
	\begin{itemize}
		\item $t_1$= (\textbf{\texttt{x=[0], y=0}}),
		\item $t_2$= (\textbf{\texttt{x=[3,5,5,3], y=4}}),
		\item $t_3$= (\textbf{\texttt{x=[1,1,1], y=2}}).
	\end{itemize}
	satisfies branch coverage and reveals all three bugs, since $t_2$ detects 
	Bug~1 and Bug~3, and $t_3$ detects Bug~1 and Bug~2.
	In contrast, the test suite 
	\begin{itemize}
		\item $t_1$= (\textbf{\texttt{x=[0], y=0}}),
		\item $t_5$= (\textbf{\texttt{x=[3,1,1,2], y=1}}).
	\end{itemize}
	also satisfies branch coverage, but reveals none of the three bugs, since $t_1$ 
	is unable to reach any bug and the execution of $t_5$ (accidentally) produces a correct output 
	for all versions. 
	These examples illustrate the well-known dilemma of unit-test selection that
	(1) adherence to (purely syntactic) code-coverage criteria does 
	not guarantee effective test-suite selections~\cite{Pecorelli2021, Inozemtseva2014}, and
	(2) using more test cases does not necessary 
	increase effectiveness, yet obviously decreases testing efficiency.
	Concerning (2), \emph{TSR} aims at 
	removing redundant test cases from a test suite without decreasing code coverage~\cite{Harrold1993}.

	\paragraph{Test-Suite Reduction.}
	Let us now assume that a developer/tester first selects the 
	test cases $t_1$ and $t_2$ for testing program version $P_0$, 
	where $t_2$ fails due to Bugs 1 and 3.
	After applying the first bug fix to remove Bug 1, the initial test suite
	consisting of $t_1$ and $t_2$ does no more satisfy branch coverage. 
	This is due to the fact, that after the first bug fix is applied, 
	the first element (i.e., the meta-information) of the input array is no more included into the search. 
	As the last element is also not included (due to Bug~2), execution of $t_2$ will not enter the 
	\textbf{\texttt{true}}-branch of the \textbf{\texttt{if}}-statement in line~\ref{line:if-bug_p1}, and is, therefore, 
	unable to reach 100\% branch coverage.
	Hence, the developer/tester has to select a further test case, for instance:
	\begin{itemize}
		\item $t_6$= (\textbf{\texttt{x=[3,0,1,0], y=0}})
	\end{itemize}
	to cover the missing branch.
	Thus, the existing test case $t_2$ becomes \emph{redundant} and
	might be removed from the test suite as $t_1$ and $t_6$ are sufficient for
	100\% branch coverage. 

	More generally, \emph{TSR} is concerned with
	selecting from an existing set of test cases a sufficient, yet preferably small number
	of test cases for a program under test.
	Corresponding strategies for TSR have to address several
	challenges:
	\begin{itemize}
		\item Finding a \emph{minimal} set of test cases from
		an existing test suite satisfying a given coverage criterion 
		is NP-hard, being reducible to the minimum set-cover problem~\cite{Khalilian2012}.
		\item As illustrated by the running example, defining redundancy of test cases only with respect 
		to a code-coverage criterion might be misleading thus obstructing
		testing effectiveness~\cite{Pecorelli2021, Inozemtseva2014}.
	\end{itemize}
	A further challenge arises in the context of evolving programs.
	Let us next assume that bug fix 2 is applied to remove Bug 2. 
	After that, the new test case $t_6$ is no more able to reveal the remaining Bug 3, as the last element 
	of the array is equal to $y$ and, therefore, the test output (accidentally) satisfies the specification.
	In contrast, the previously removed test case $t_2$ would have revealed Bug 3, since the last 
	element is smaller than $y$, thus leading to the output of $3$ instead of $-2$.
	This example shows that TSR solely based on structural coverage criteria 
	defined on a current program version might be problematic in case of evolving programs. 
	Test cases being redundant for a current program version 
	might become relevant again after a program revision.
	In particular, \emph{regression testing}\footnote{The term regression testing often summarizes 
		a wide range of different disciplines of testing evolving programs 
		not only including RTS, but also
		test-case prioritization, test-history analysis, test-artifact storage etc.
		We will focus on to the core problem of selecting test cases 
		for regression testing.} is concerned with selecting a suitable set of test cases
	after program revisions~\cite{Yoo2012, Kim2002}.

	\paragraph{Regression Testing.}
	As illustrated by the previous example, after revealing a bug due to a failed test-case execution,
	a developer/tester consecutively creates program revisions to fix the bugs
	\footnote{For the sake of clarity, we assume in the following that the granularity of each individual modification made 
		during a revision is limited to one particular line of code.
		However, this assumption does not possess any threat to the validity of our approach.}.
	After creating a program revision, the current test suite is re-executed to
	assure modifications made during the revision (1) successfully fix a discovered bug 
	(i.e., test cases previously failing now pass) and (2) do not introduce new bugs
	(i.e., test cases previously passing still pass).
	Those test cases from the existing test suite should be re-executed 
	investigating program parts affected by modifications made during the revision.
	New test cases might be required to investigate existing and/or newly introduced program parts 
	not yet and/or no more covered by existing test cases.
	In both cases, one distinguishes between \emph{modification-traversing} 
	test cases and \emph{modification-revealing} test cases~\cite{Yoo2012}.
	Executions of modification-traversing test cases at least \emph{reach} 
	a program modification, but may, however,
	not reveal the modification to a tester (i.e., the program versions before and after the revision may produce the same output values for the same test inputs).
	In contrast, executions of modification-revealing test cases not only reach a program modification, 
	but also yield different output values when applied before and after the revision.

	For instance, recall the
	case in which the initial test suite consisting of $t_1$ and $t_2$ 
	is applied to $P_0$.
	As $t_2$ fails on $P_0$, a developer might perform the first bug fix, leading to $P_1$.
	Next assume that $t_1$ and $t_2$ are simply reused for testing $P_1$.
	Both test cases would pass now although the second and third bug are still
	present in $P_1$. 
	This is because $t_1$ is unable to reach the bug 
	and $t_2$ is unable to detect the bug as the last element which 
	would reveal Bug~3 is not included in the search due to Bug~2. 
	The output yielded by $t_2$, therefore, also conforms to the specification.
	In contrast, selecting $t_3$ in addition to, or instead of, $t_2$ 
	would also reveal the second bug on $P_1$, thus enabling
	the developer to perform a further bug fix leading to $P_2$.
	However, after this bug fix, $t_3$ becomes unable to detect the last 
	bug as there are no values contained in $x$ having a smaller value than $y$.
	In contrast, $t_2$ is now able to detect the last bug.

	The previous example shows that effectiveness 
	of a test suite not only depends on the particular test cases, but also on the current program version under test. 
	Hence, fault-detection probability of test cases may both decay as well as improve over time.

	To generalize, \emph{RTS} is concerned with
	selecting a sufficient, yet preferably small number of existing/new test cases 
	to be (re-)executed on a revised program.
	Strategies for RTS have to address several
	challenges:
	\begin{itemize}
		\item Finding a \emph{minimal} modification-traversing set of test cases from
		an existing test suite is NP-hard and does, by definition, not guarantee effective assurance of program revisions.
		\item Finding \emph{any} modification-revealing test case for a program modification 
		corresponds to the program-equivalence problem which is undecidable.
		\item In a practical setting, the aforementioned idealized assumptions on
		program revisions do usually not hold: units may contain several
		bugs which influence or even obfuscate each other and not all modifications
		applied during a program revision are actually bug fixes.	
	\end{itemize}

	\paragraph{Test-Suite Reduction vs. Regression-Test Selection}
	
	Both techniques aim at \emph{improving testing efficiency} by 
	reducing the number of test cases to be executed without
	presumably harming effectiveness:
	\begin{itemize}
		\item TSR strategies aim at selecting from an existing test suite
		of one program version 
		a presumably small subset being sufficient to satisfy the given code-coverage criterion.
		\item RTS strategies aim at selecting from an existing test suite of previous
		program versions and/or new test cases a presumably small subset being sufficient to traverse/reveal
		critical program modifications in the current program revision.
	\end{itemize}
	The previous examples illustrate the subtle interplay 
	between both strategies affecting efficiency and effectiveness in non-obvious ways:
	\begin{itemize}
		\item Removing too few test cases during TSR
		might obstruct the gain in efficiency of subsequent RTS as well as TSR steps
		due to critically increasing overhead required for those strategies in case of ever-growing test suites.
		\item Removing too many test cases during TSR
		might obstruct effectiveness of subsequent RTS as well as TSR steps
		as currently redundant test cases might become relevant again.
	\end{itemize}

	We next present a novel evaluation framework to systematically investigate these interactions.

	\section{Evaluation Methodology}\label{sec3:meth}
	We present a configurable framework for 
	systematically addressing the challenges described in the previous section.
	The methodology allows us to evaluate 
	interactions between different strategies for TSR and RTS
	with respect to practical efficiency and effectiveness measures.
	We first present a conceptual overview and then describe 
	the provided parameters for adjusting the strategies.

	\subsection{Overview}
\begin{figure}[tp]
		\centering
		\includegraphics[width=\linewidth]{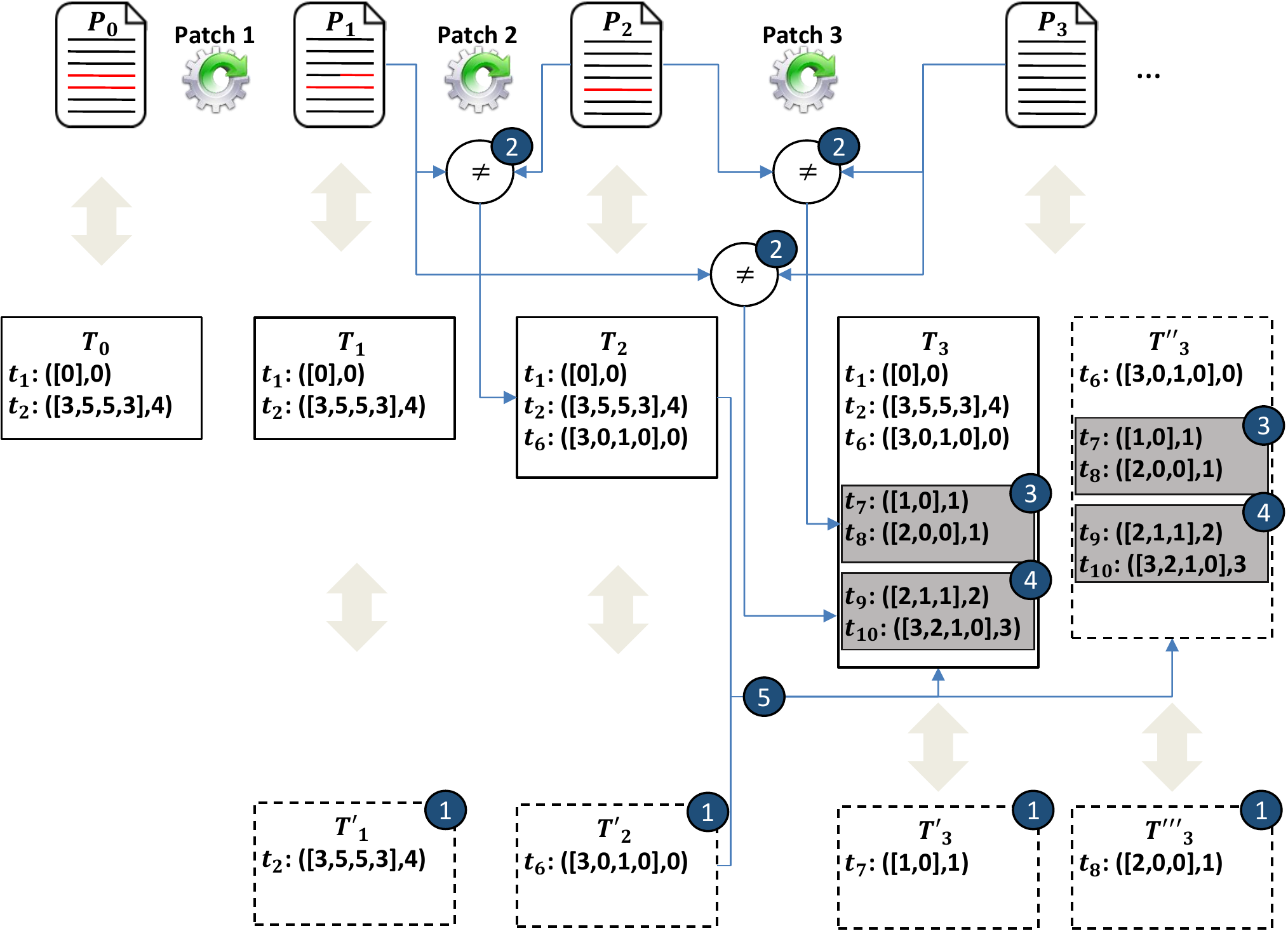}
		\caption{Overview of the Evaluation Methodology}\label{fig:bigp}
	\end{figure}

	Figure~\ref{fig:bigp} provides an overview of our methodology
	using the running example introduced in the previous section.
	
	Starting from the initial program version $P_0$, the subsequent 
	program versions, $P_1, P_2,\allowbreak P_3, \ldots$, result from applying
	consecutive revisions, given as patches $\textit{Patch}_{1}$, $\textit{Patch}_{2}$, $\textit{Patch}_{3}$.
	For each program version $P_i$, we consider a corresponding
	test suite $T_i$ containing the test cases selected for this particular program version.
	Based on the initial test suite $T_0$ created for $P_0$ (e.g., using either a coverage 
	criterion like branch coverage, or randomly generated test cases, or other criteria), the 
	regression test suites $T_i$ for testing subsequent program versions $P_i$, $i>0$, result from
	applying a RTS strategy.
	As a complementing step, a TSR strategy may be applied
	to test suites $T_i$ of program versions $P_i$ to remove redundant test cases.

	Our methodology comprises strategies
	for both \emph{TSR} as well as \emph{RTS}.
	The parameters for fine-tuning the strategies are denoted as circled numbers
	in Fig.~\ref{fig:bigp}.
	We briefly describe these parameters which will 
	be explained in more detail below.
	\begin{enumerate}[label=\protect\raisebox{.5pt}{\textcircled{\raisebox{-.5pt} {\small \arabic*}}}]
		\item \textbf{Reduction Strategy (RS):} Technique used for TSR. 
		Possible strategies: \textbf{None}, \textbf{ILP}, \textbf{FAST++}, and \textbf{DIFF}.
		
		\item \textbf{Regression Test-Case Selection Criterion (RTC):} 
		New regression test cases for a program version $P_i$ 
		may be added to test suite $T_i$ either by means 
		of modification-traversing test cases (i.e., at least
		reaching the lines of code modified from $P_{i-1}$ to $P_{i}$) 
		or by modification-revealing test cases (i.e., yielding different 
		outputs if applied to $P_{i}$ and $P_{i-1}$).
		\item \textbf{Number of Regression Test Cases (NRT):}
		The number of \emph{different} regression test cases
		added into $T_i$ satisfying \textbf{RTC} for \emph{each} previous
		program version $T_j$, $0\leq j < i$.
		\item \textbf{Number of Previous Program Revisions (NPR):} 
		The (maximum) number of previous program 
		versions $P_j$ ($i-\textbf{NPR}\leq j < i$) for all of which
		\textbf{NRT} different test cases satisfying \textbf{RTC} are added to $T_i$.
		
		\item \textbf{Continuous Reduction (CR):} 
		Controls whether the 
		non-reduced test suite $T_{i-1}$ (\textbf{No-CR}) or the reduced test suite 
		$T_{i-1}'$ (\textbf{CR}) of the previous program $P_{i-1}$ version is (re-)used 
		for the next program version $P_{i}$ or if the previous test cases are ignored (\textbf{None}).
		This test suite is extended by new test cases according to
		the previously described parameters.
	\end{enumerate}
	Note that parameter \textbf{CR} can be applied for all test suites of all versions. 
	However, for clarity and space reasons, it is only depicted from
	program version $P_2$ on in Fig.~\ref{fig:bigp}.

	\subsection{Test-Suite Reduction Strategies}\label{sec:reduction-strategies}
	We now describe the different TSR
	strategies supported by our framework.
	We, again, use an illustrative example to explain
	the impact of the strategies on the trade-off between
	precision and computational effort.
	We first give a general characterization
	of the \emph{test-suite minimization problem} for 
	one single program $P$, a given test suite $T$
	and a code-coverage criterion on $P$~\cite{Harrold1993}.

	\bigskip
	\noindent\textbf{Input:} Program $P$, Test Suite $T$, where
	\begin{itemize}
		\item $P$ contains of a set of test goals $G=\{g_1, g_2,\ldots, g_n\}$ 
		according to the given code-coverage criterion and
		\item test suite $T={t_1, t_2,\ldots, t_n}$ consists of a set of test cases, 
		where each $t_j\in T$ covers a subset $G_j\subseteq G$ of test goals on $P$ such that
		for each $g_i\in G$ there exists at least one test case $t_j\in T$ with
		$g_i \in G_j$~\footnote{Set $G$ actually contains the subsets of test goals
			covered by at least one test case in $T$.}.
	\end{itemize}
	\noindent\textbf{Output:} Minimal Test Suite $T'\subseteq T$, where
	\begin{itemize}
		\item for each $g_i\in G$ there exists at least one test case $t_j\in T'$ with
		$g_i \in G_j$ and
		\item for each $T''\subseteq T$ also satisfying the first property,
		it holds that $|T'|\leq |T''|$.
	\end{itemize}
	The test-suite minimization problem is NP-hard being reducible to the minimum set cover problem~\cite{Harrold1993}
	such that finding exact minimal solutions is computational infeasible
	for realistic programs.
	Various \emph{TSR} heuristics 
	have been proposed for approximating minimal test suites 
	constituting different trade-offs
	between precision (deviation from exact solutions) and
	computational effort for performing the reduction.

\begin{figure}[tp]
		\centering
		\includegraphics[width=\linewidth]{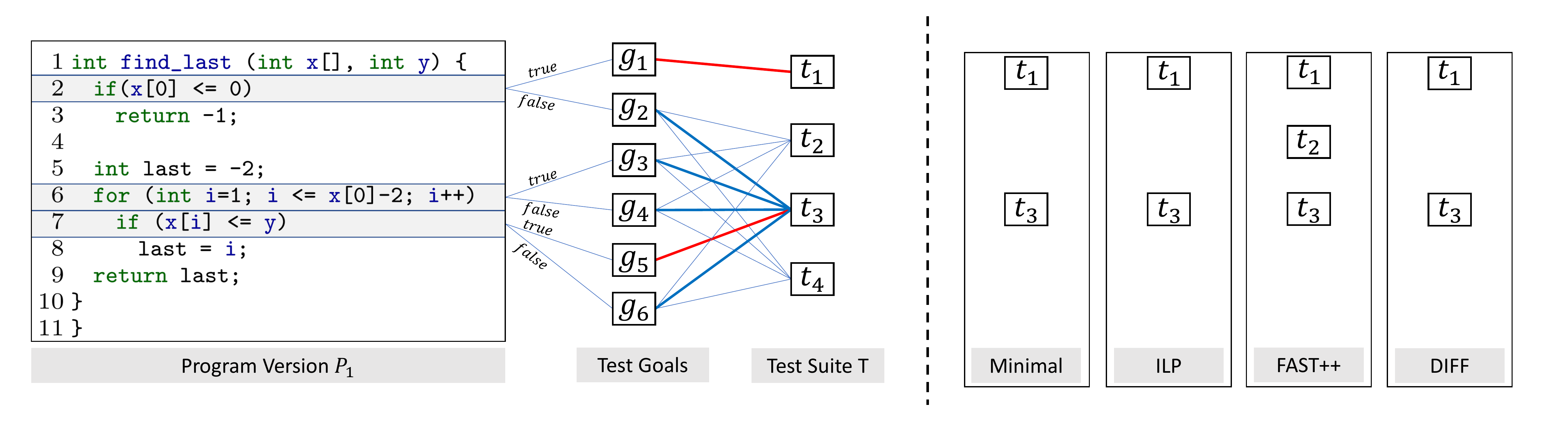}
		\caption{Comparision of Test-Suite Reduction Strategies}\label{fig:TSReduction}
	\end{figure}

	To illustrate the different approaches in our 
	framework, $P_1$ from our running example together with 
	$\textit{T}_3$ containing four test cases selected for branch coverage:
	\begin{itemize}
		\item $t_1$= (\textbf{\texttt{x=[0], y=0}}),
		\item $t_2$= (\textbf{\texttt{x=[3,5,5,3], y=4}}),
		\item $t_3$= (\textbf{\texttt{x=[1,1,1], y=2}}), and
		\item $t_4$= (\textbf{\texttt{x=[1,2,2], y=0}}).
	\end{itemize}
	Figure~\ref{fig:TSReduction} (on the left) illustrates the test-suite minimization problem.
	Program version $P_1$ contains three conditional branches (i.e.,
	if-statement in line 2, loop-head in line 6 and if-statement in line 7), 
	leading to six goals $G=\{g_1, g_2, \ldots, g_6\}$.
	A line between test case $t_i$ and test goal $g_j$ indicates
	that $g_j \in G_i$ (i.e., $t_i$ satisfies $g_j$).
	Test suite $T_3$ indeed satisfies branch coverage on $P_1$ as for each
	$g\in G$, there is at least one test case in $T_3$.
	However, $T_3$ is not \emph{minimal} as, for instance, $t_2$ and $t_4$ 
	cover exactly the same test goals, $g_2, g_3, g_4$ and $g_6$, and this set is
	further subsumed by the test goals satisfied by $t_3$.
	In fact, $t_3$ is the only test case covering $g_5$ and 
	$t_1$ is also indispensable as it is the only test case satisfying $g_1$.
	Hence, $T_3 = \{t_1, t_3\}$ is the (unique) \emph{minimal} test suite as shown on the left.

	Computing the minimal test suite requires, in the worst case, to enumerate all $2^{|T|}$
	subsets of $T$ which is infeasible in case of realistic programs.
	Practical approaches usually compute \emph{reduced} test suites approximating
	the minimal solution.
	We now describe the strategies considered in our framework.

	\paragraph{ILP Strategy.}
	The test-suite minimization problem can be encoded as 
	Integer linear optimization problem which can be precisely
	solved using Integer Linear Programming (ILP) solvers~\cite{Khalilian2012}.
	Our ILP encoding uses for each test case $t_i\in T$
	a decision variable $x_i$ either having value $1$ if
	$t_i$ is selected, or value $0$ if $t_i$ is not selected.
	The ILP formula contains for each test goal $g_j\in G$ a clause
	building the sum of the decision variables $x_i$ of all test cases
	$t_i\in T$ for which $g_j\in G_i$ holds.
	By requiring the value of each such sum to be greater than $0$, we ensure
	each test goal to be covered by the minimal test suite.
	To enforce minimality, the overall optimization objective is to minimize
	the sum over all values of decision variables $x_i$.

	Applied to our example, this encoding introduces
	the variables $x_1, x_2, x_3, x_3$ for the test cases
	in $T$ and adds the following clauses for the test goals:
	\begin{align}
		x_1 							& >= 1 \\
		x_2 + x_3 + x_4		& >= 1 \\
		x_2 + x_3 + x_4		& >= 1 \\
		x_2 + x_3 + x_4		& >= 1 \\
		x_3								& >= 1 \\
		x_2 + x_3 + x_4		& >= 1 \\
	\end{align}
	The optimization objectives is defined as:
	$$\textit{min}(x_1 + x_2 + x_3 + x_4)\text{.}$$
	As illustrated in Fig.\ref{fig:TSReduction}, recent ILP solvers
	are able to deliver an exact minimal solution. However, large amount of test cases naturally lead to
	high computational effort.
	We discuss two TSR heuristics 
	to avoid intractable computational effort in case of 
	larger-scaled problems, yet providing an acceptable approximation
	of the exact (minimal) solution.

	\paragraph{FAST++ Strategy.}
	The FAST++ approach encodes the selection of test cases into a (reduced) test suite 
	using the Euclidean distance of a reduced vector-space model 
	computed by random projection~\cite{Cruciani2019}.
	We also illustrate this approach using our example.
	Test cases are encoded as vectors, where the number of elements 
	corresponds to the number of different input values of all test cases. 
	The first element of each vector denotes the number of 
	occurrences of the lowest input value (e.g., for our running example the number '0'). 
	The next element of each vector is the next lowest input value (e.g., for our running example the number '1').
	The encoding of all test cases is shown in Tab.~\ref{tab:tc_vectorsRP} and 
	the result of reducing the dimensions of the vector-space model 
	by random projection is shown in Tab.~\ref{tab:tc_vectors}.
	
	\begin{table}[pt]
		\caption{Vector Encoding of Test Cases}\label{tab:tc_vectorsRP}
		\centering
		\begin{tabular}{|l|lccccc|}
			\hline
			Test Case \textbackslash~Value & 0 & 1 & 2 & 3 & 5 & 6 \\
			\hline
			($t_1$)   & 2  & 0  & 0 & 0 & 0 & 0 \\
			($t_2$)   & 0  & 0  & 0 & 2 & 1 & 2 \\
			($t_3$)   & 0  & 3  & 1 & 0 & 0 & 0 \\
			($t_4$)   & 1  & 1  & 2 & 0 & 0 & 0 \\
			
			\hline
		\end{tabular}
	\end{table}

	\begin{table}[pt]
		\caption{Vector Encoding of Test Cases after Random Projection}\label{tab:tc_vectors}
		\centering
		\begin{tabular}{|l|lcc|}
			\hline
			Test Case \textbackslash~Value& x & y & z   \\
			\hline
			($t_1$)   & 0  & 0  & 0 \\
			($t_2$)   & 0  & 0  & 5 \\
			($t_3$)   & 0  & 3  & 2 \\
			($t_4$)   & 0  & 2  & 0 \\
			\hline
		\end{tabular}
	\end{table}
	
	Based on this encoding, TSR works as follows.
	First, a random test case (e.g., $t_2$) is selected
	into the (initially empty) reduced test suite.
	For the selected test case, the Euclidean distances to all other remaining
	test cases is computed based on the reduced vector-space model.
	The probability of selecting a particular remaining test case next 
	into the reduced test suite increases with the distance value 
	(thus preferring test cases covering test goals being dissimilar
	to those covered by previously selected ones).
	In our example, this would be, for instance, test case $t_1$.
	This iterative step is repeated until all test goals are covered.
	For instance, the next test case might be $t_3$ which suffices
	to achieve full branch coverage thus leading to termination.
	This technique is, on average, more efficient than ILP, yet potentially leading to less precise results
	as demonstrated by the example.
	There is not even a guarantee to find local optima  
	as random projection may obfuscate necessary information. 
	Additionally, the ordering of input values is ignored. 

	\paragraph{DIFF Strategy.}
	This strategy also incrementally 
	selects test cases until all test goals are covered~\cite{TestCov}.
	In contrast to FAST++, DIFF is a purely greedy-based approach which 
	always selects as next test case one that covers a maximum number
	of uncovered test goals.
	Applied to our example, DIFF would, for instance, perform the following selections:
	\begin{enumerate}
		\item $t_3$ (covering 5 uncovered test goals),
		\item $t_1$ (covering 1 uncovered test goal).
	\end{enumerate}
	This technique is, in general, similarly efficient and (im-)precise as FAST++, where
	this particular example is well-suited for DIFF as the local
	optimum is the same as the global one. 
	However, if local and global optima differ, 
	there is no guarantee about the optimality of the result.
	
	The different TSR technique might lead to different results in terms of the number of test cases of the reduced test suite.
	Although, this clearly leads to different results in terms of efficiency (in terms of test-suite size) it also might affect effectiveness.
	For example, test case $t_2$ does not cover additional test goals and also does not detect any bugs in program version $P_1$. However, in program version $P_2$ it will again detect Bug~3.
	We chose those three TSR strategies as they provide different trade-offs between the effectiveness of reduction (i.e., how small the resulting test-suite will be) and efficiency in terms of CPU time needed for reduction.~\cite{Cruciani2019, DegradedILP2008} Since ILP provides an optimal solution, the resulting test-suite will be minimal, however, leading to much more computational effort. The greedy approach (DIFF) is more efficient, however, the resulting test-suite might be larger compared to ILP. FAST++ is even more efficient compared to the greedy approach, however, leading to to even larger test-suites.

	\subsection{Regression-Test Selection Strategies}
	The TSR strategies considered
	so far are concerned with selecting from an \emph{existing test suite
		of one single program version} a reduced subset preserving
	test coverage on that program version.
	In contrast, RTS strategies are concerned
	with selecting from an \emph{existing 
		test suite of a previous program version} a sufficient set of test cases to investigate
	modifications applied to that program version leading to a subsequent version.
	If no such test case(s) can be found among the existing ones, new test cases must be created
	and added to the test suite of the subsequent program version.
	We next describe different RTS
	strategies supported by our framework.
	We use an illustrative example to explain
	the impact of the strategies on the trade-offs between
	precision and computational effort.
	We start with a general characterization
	of the \emph{regression-test selection problem}, consisting of 
	a program version $P$ with existing test suite $T$
	and subsequent version $P'$ for which a regression test suite $T'$
	is created~\cite{Yoo2012}.

	\bigskip
	\noindent\textbf{Input:} Program version $P_{i-1}$ with Test Suite $T_{i-1}$ and subsequent program version $P_{i}$.
	
	\noindent\textbf{Output:} Regression Test Suite $T_{i} = T_{i-1}' \cup T_{i}'$, where
	$T_{i-1}'\subseteq T_{i-1}$ are existing test cases selected for reuse from $T_{i-1}$
	and $T_{i}'$ are newly added test cases.
	\bigskip
	
\begin{figure}[tp]
		\centering
		\includegraphics[width=\linewidth]{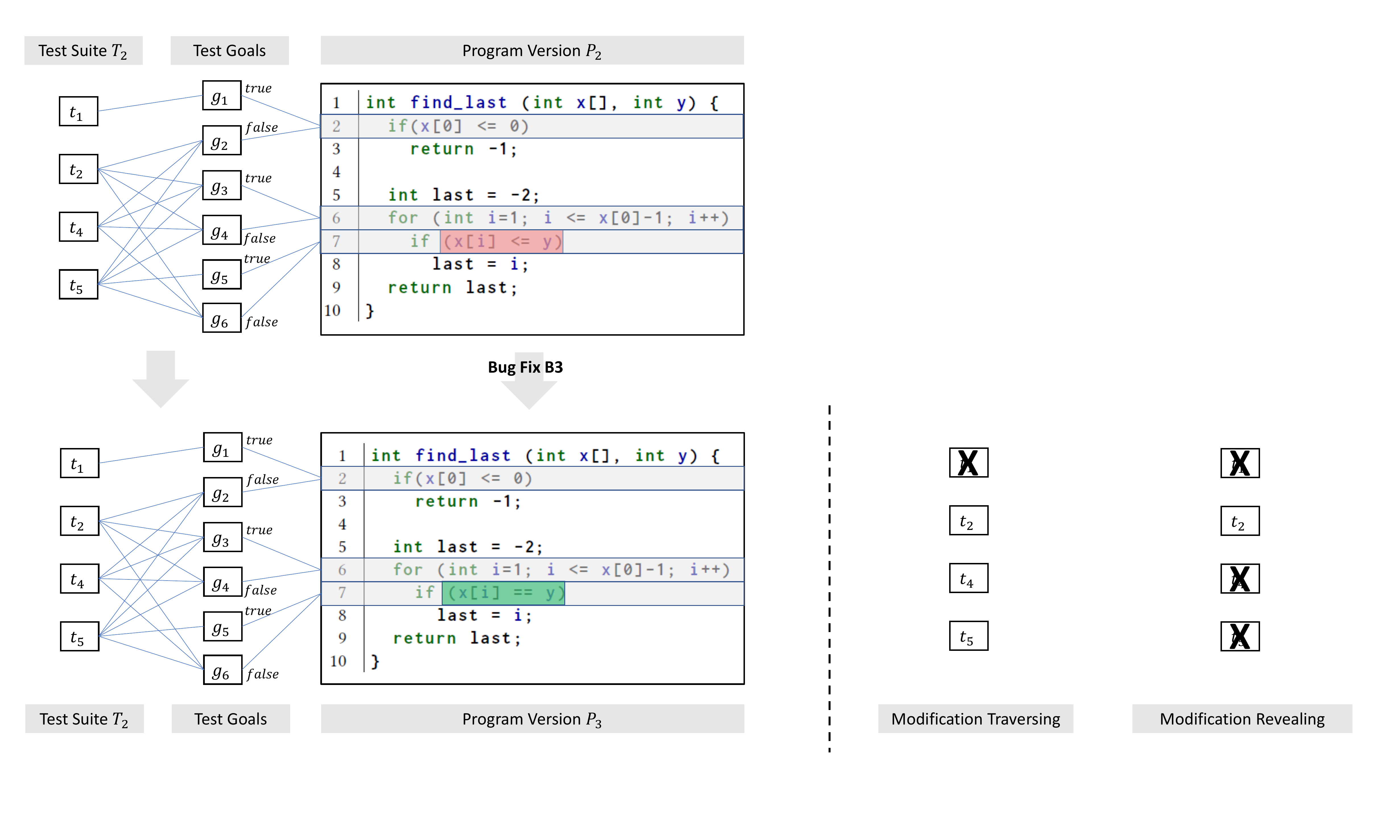}
		\caption{Comparison of Regression-Test Selection Strategies}\label{fig:RTSelection}
	\end{figure}

	
	Consider $P_3$
	from our running example together with test suite $\textit{T}_3$.
	Figure~\ref{fig:RTSelection} provides an illustration of the 
	RTS problem.
	For $P_3$, the selected test suite contains three 
	modification-traversing test cases (i.e., $t_2$, $t_3$ and $t_4$) being 
	able to at least \emph{traverse} the modified line of code. 
	However, only $t_2$ is \emph{modification revealing} as it not only reaches
	the modification, but also disposes 
	a difference in the output behavior of $P_3$ as compared to $P_2$.
	Hence, depending on the RTS strategy,
	also new test cases have to be generated for $T_{i}'$.
	
	\paragraph{Regression-Test-Case Generation.}
	
	We now describe how we automatically generate regression test cases
	utilizing off-the-shelf test-generation tools originally designed for covering 
	test goals encoded as reachability properties.
	We utilize a software model-checker for C programs
	to automatically derive test cases from counterexamples
	for (non-) reachability queries of test goals~\cite{Beyer2004}.
	We encode the problem of generating regression test cases for different versions of a program 
	in a generic \texttt{comparator}-function.
	The shape of the comparator-function $P_{i,j}$ for the program versions $P_i$ and $P_j$ 
	depends on parameter $\textbf{RTC}$. 
	We illustrate this using the comparator-function $P_{2,3}$ for our running example.

	\begin{figure}[tp]
		\subfloat[Modification-Traversing Comparator-Function for $P_2$ and $P_3$]{\label{fig:compare1}\parbox{\linewidth}{\centering\usebox{\compareMT}}}
		
		\subfloat[Modification-Revealing Comparator-Function for $P_2$ and $P_3$]{\label{fig:compare2}\parbox{\linewidth}{\centering\usebox{\compareMR}}}
		\caption{Comparator-Functions}
	\end{figure}
	
	\begin{itemize}
		\item $\textbf{RTC}=\textbf{MT}$: 
		The comparator-function $P_{2,3}$ for generating \emph{modification-traversing} test cases
		between $P_2$ and $P_3$ is shown in Figure~\ref{fig:compare1}.
		The program location affected by the program modification 
		leading from $P_2$ to $P_3$ is marked by a special program label $\textit{test\_goal}$.
		A test-case generator can be used to generate a program input reaching
		this program label (and therefore at least \emph{traversing} this modification). 
		
		\item $\textbf{RTC}=\textbf{MR}$:
		The comparator-program $P_{2,3}$ for generating \emph{modification-revealing} test cases
		between $P_2$ and $P_3$ is shown in Fig.~\ref{fig:compare2}. 
		Here, the program input is delegated to $P_2$ and $P_3$ and the output is 
		compared by an \textit{equals}-function (or \textbf{\texttt{!=}} in case of basic types) based on the return-type.
		Hence, the special program label $\textit{test\_goal}$ in line 3 is only reachable
		by a test case if the test input generated by a test-case generator 
		yields different output values for $P_2$ and $P_3$ (thus not only reaching, but also 
		\emph{revealing} this modification).	
	\end{itemize}
	
	\paragraph{Remarks.}
	The problem of generating modification-revealing test cases obviously subsumes
	the generation of modification-traversing test cases and therefore, presumably, causes
	(much) more computational effort.
	In fact, \emph{both} problems are undecidable as reachability of 
	program locations is reducible to the halting problem.
	Hence, the test-generation approach described above is inherently incomplete~\footnote{
		This fact, however, is already true for the problem of generating a test case simply reaching
		a test goal in one version of a program in a non-regression scenario.}.
	Our framework further comprises the parameter \textbf{NRT} to control 
	the number of \emph{multiple different test cases} distinguishing the output behavior of 
	two consecutive program versions (which may, however, fail due to the theoretical limitations described before). 
	Lastly, the technique based on the comparator-function suffers from some technical limitations. 
	For example, if the data type of the return value changes the comparison 
	of the return values is, in general, no more possible (see Fig.~\ref{fig:compare_invalid}, 
	where the return types of function $find\_last\_p3$ and $find\_last\_p4$ differ
	such that the comparator-program would not compile). 
	Additionally, as both program versions need to be compiled into one 
	single program, merging of those versions can become technical challenging (e.g., 
	if the data type of global variables or functions changes or the members of structs are modified).
	\begin{figure}[tp]
		\parbox{\linewidth}{\centering\usebox{\compareInvalid}}
		\caption{Invalid Comparator Program}\label{fig:compare_invalid}
	\end{figure}

\begin{figure}[tp]
		\hfill
		\subfloat[Original CFA]{\label{fig:cfa}\includegraphics[width=.3\linewidth]{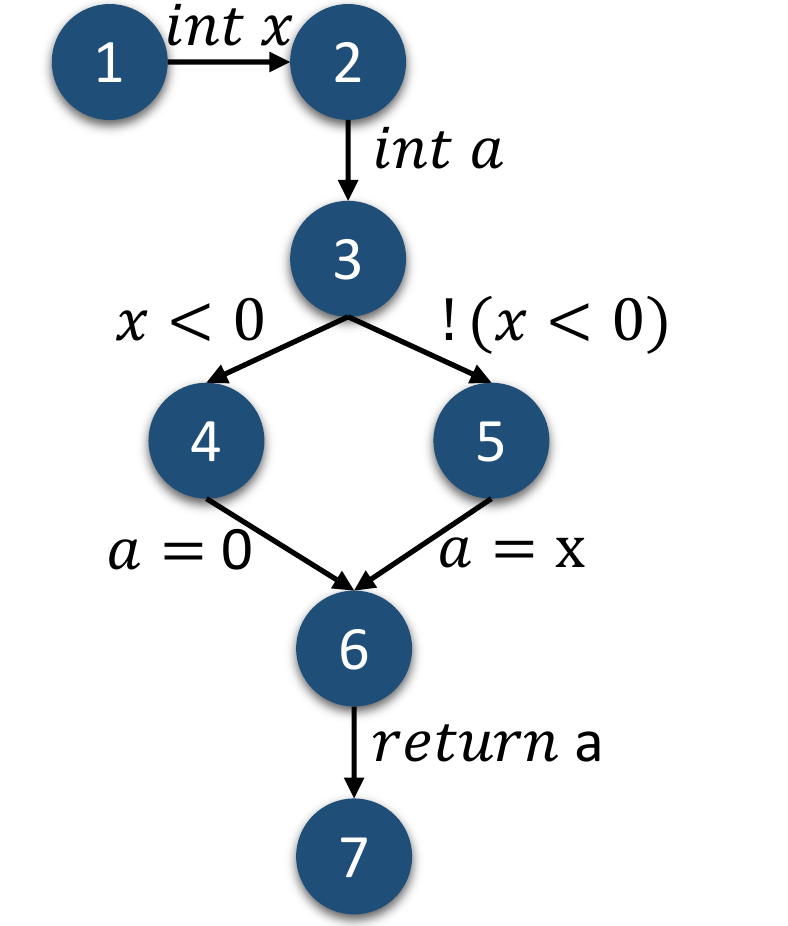}}
		\hfill
		\subfloat[Instrumented CFA]{\label{fig:cfa_weaved}\includegraphics[width=.3\linewidth]{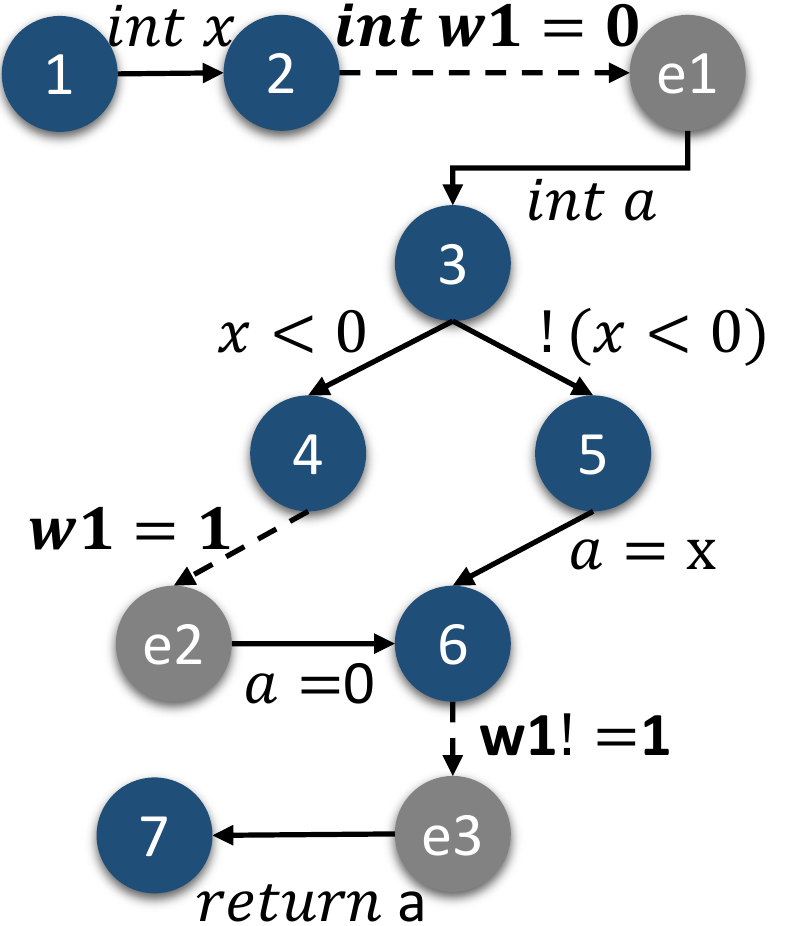}}
		\hfill\strut
		\caption{On-the-fly CFA Instrumentation}\label{fig:Weaving}
	\end{figure}

	\paragraph{Multiple Regression Test-Case Generation.}\label{subsec:multiple-regression-tc-gen}
	Most test-case generators transform the input program into an intermediate 
	representation like a control-flow automaton (CFA).
	To keep the following examples graspable, we use the basic 
	example shown in Figure~\ref{fig:cfa} instead of our running example. 
	This CFA corresponds to a small program with input value \textbf{\texttt{x}} returning 
	\textbf{\texttt{0}} if \textbf{\texttt{x}} is smaller than zero and the value of \textbf{\texttt{x}}, otherwise.
	A CFA can be used to identify all program paths leading to test goals 
	(e.g., all CFA-edges in case of branch coverage) and
	to control automated test-suite generation by performing reachability queries 
	for uncovered test goal~\cite{Beyer2013}.
	To support parameter $\textbf{NRT}$, we have to generate \emph{multiple different} 
	test cases reaching the same test goal (i.e., a modified program location).
	We start by generating the first test case based on the original CFA. 
	For instance, to generate a test case reaching 
	the \textbf{\texttt{return}}-edge in Fig.~\ref{fig:cfa}, any input is feasible (e.g., input \textbf{\texttt{-5}} 
	would traverse the left path).
	For further test cases reaching the \textbf{\texttt{return}}-edge, 
	we modify the CFA as shown in Fig.~\ref{fig:cfa_weaved} to exclude already covered paths.
	For this, we add for each path for which we already created a test case a 
	fresh variable with initial value \textbf{\texttt{0}} (i.e., variable \textit{w1} introduced by the new edge 
	from node \textit{2} to \textit{e1}).
	We assign value \textbf{\texttt{1}} to this variable after traversing the associated branch of the path
	(i.e., the new edge from node \textit{4} to \textit{e2}). 
	By further requiring the value of at least one of the fresh variables before 
	reaching the test goal must not be \textit{1} (i.e., the new edge from node \textit{6} to \textit{e3}), 
	we exclude all previous test cases reaching that test goal.
	In the example, assumption $\textbf{\texttt{w1!=1}}$ enforces 
	a further test case to assign a positive value to \textit{x}, whereas a
	third run of the test-case generator will find no further path
	reaching this goal.
	
	For example, to generate two test cases reaching line~\ref{line:return-i_p3} 
	in program version $P_3$, we first generate a test case as usual (e.g., $t_7$= (\textbf{\texttt{x=[1,0], y=1}})). 
	For the second test case, we introduce a fresh variable into the CFA which is incremented when 
	traversing a branch of the path already taken by the previous test case (i.e., the \textbf{\texttt{true}}-branch of 
	the \textbf{\texttt{if}}-statement, the \textbf{\texttt{true}}-branch of the \textbf{\texttt{for}}-loop, 
	the \textbf{\texttt{false}}-branch of the \textbf{\texttt{for}}-loop and the \textbf{\texttt{false}}-branch 
	of the last \textbf{\texttt{if}}-statement).
	As there are four branches, the final check will ascertain that the fresh variable is not equal to $4$.
	When iterating through the \textbf{\texttt{for}}-loop more than once (as opposed to 
	the first test case) or taking the \textbf{\texttt{true}}-branch of the last \textbf{\texttt{if}}-statement, 
	the value will differ from $4$ and, therefore, the path can be taken for 
	a further test case being different from all previous ones (e.g., $t_8$= (\textbf{\texttt{x=[2,0,0], y=1)}}).
	
	\paragraph{Remarks.} Again, this methodology is \emph{inherently incomplete} as it is unknown 
	whether further test cases exist if no more test cases are found by
	the test-case generator (e.g., due to time-outs).
	In addition, each incrementally added fresh variable presumably
	increases program complexity thus potentially 
	increasing the computational effort required for every further test case.

	\paragraph{Regression Test-Suite Generation.}
	We conclude by summarizing the description of our framework. 
	Algorithm~\ref{algorithm:tcgen} depicts a 
	generic procedure for generating regression test suites $T_i$ 
	for program version $P_i$ meeting all possible 
	instantiations of the parameters $\textbf{RTC}$, $\textbf{NRT}$, $\textbf{NPR}$, $\textbf{RS}$ and $\textbf{CR}$.
	\begin{figure}[tp]
		\centering
	\begin{minipage}{0.85\linewidth}
	\begin{algorithm}[H]

		\begin{algorithmic}[1]
			\footnotesize
			\Require{$\textbf{RTC}\in\{\textit{MT},\textit{MR}\}$, $\textbf{NRT}\in\mathbb{N}^+$, $\textbf{NPR}\in\{1,\ldots,i\}$}
			\Require{$P_i$, $\textit{Patch}_1, \ldots, \textit{Patch}_i$}
			\Require{$T_{i-1}$}
			\Ensure{$T_i$}
			\State \textbf{procedure}\textsc{Main}{} \label{line:algo-main-head}
			\State $T_i \leftarrow T_{i-1}(\textbf{CR})$ \label{line:algo-Ti-initial}
			\State $T_{i,j} \leftarrow \emptyset$ \label{line:algo-tij-initial}
			\State $P_j \leftarrow P_i$ \label{line:algo-Pj-initial}
			\For{$k = 0$ to $\textbf{NPR}-1$} \label{line:algo-for-loop-head}
			\State $P_j \leftarrow \textsc{apply}~\textit{Patch}_{i-k}^{-1}$ to $P_j$ \label{line:algo-Pj-apply-patch}
			\State $P_{i,j} \leftarrow \textsc{Comparator}\langle P_i,P_j\rangle(\textbf{RTC})$ \label{line:algo-Pij-comparator}
			\State $l \leftarrow 0$ \label{line:algo-l0}
			\While{$l < \textbf{NRT}$} \label{line:algo-while-loop-head}
			\State $t \leftarrow \textsc{NewTestGen}(P_{i,j},T_{i,j})$  \label{line:algo-new-test-gen}
			\If{$t = \textit{null}$} \label{line:algo-if-t-null}
			\State \textit{continue in line} \ref{line:algo-for-loop-head} \label{line:algo-continue}
			\EndIf
			\State $T_{i,j} \leftarrow T_{i,j} \cup \{ t \}$ \label{line:algo-update-Tij}
			\State $l \leftarrow l+1$ \label{line:algo-lpp}
			\EndWhile
			\EndFor
			\State $T_{i} \leftarrow T_{i} \cup T_{i,j}$ \label{line:algo-update-Ti}
			\State $T_{i} \leftarrow \textsc{Reduction}\langle T_{i}\rangle(\textbf{RS})$ \label{line:algo-reduce-Ti}
		\end{algorithmic}
		
		\caption{Regression Test-Suite Generation}\label{algorithm:tcgen}
	\end{algorithm}
	\end{minipage}
\end{figure}
	Parameter $\textbf{RTC}$ is either set to
	modification-traversing (\textbf{MT}) or modification-revealing (\textbf{MR}),
	whereas $\textbf{NRT}$ and $\textbf{NPR}$ can be any number greater than zero
	($\textbf{NPR}$ is naturally restricted to be at most $i$).
	Besides the current version $P_i$, the algorithm 
	receives as inputs the sequence of patches applied since
	version $P_0$ as well as the test suite $T_{i-1}$ of
	the preceding version $P_{i-1}$ based on parameter $\textbf{CR}$. 
	Lastly, $\textbf{RS}$ is set to \textbf{ILP}, \textbf{DIFF}, \textbf{FAST++} or \textbf{None}. 

	Test suite $T_i$ is initialized with the test cases
	from the existing test suite $T_{i-1}$ (line~\ref{line:algo-Ti-initial}), 
	and $P_j$ (initialized in line~\ref{line:algo-Pj-initial} by $P_i$)
	refers to the previous version for which differentiating test cases
	are generated in the next iteration.
	The outer loop from line~\ref{line:algo-for-loop-head} to \ref{line:algo-lpp} performs a descending
	traversal over $\textbf{NPR}$ previous versions by
	reversely applying the corresponding patches to $P_j$ 
	(i.e., applying Patch$_j^{-1}$ to $P_{j}$ yields version $P_{j-1}$).
	In line~\ref{line:algo-Pij-comparator}, the \textsc{comparator}-function template is instantiated
	with $P_i$ and $P_j$ to synthesize a \emph{comparator}-function $P_{i,j}$ 
	as input for a coverage-based test-case generator in line~\ref{line:algo-new-test-gen}~\cite{Beyer2004}.
	The inner loop (lines \ref{line:algo-while-loop-head}--\ref{line:algo-lpp}) 
	repeatedly generates test cases for $P_{i,j}$ until at most
	\textbf{NRT} different test cases have been generated (cf. Sect.~\ref{subsec:multiple-regression-tc-gen}).
	The test-case generator receives as additional input the set $T_{i,j}$
	of already generated test cases.
	If the test-case generator fails to find further test cases (either due to a time-out
	or due to exhaustive search), the current iteration of the inner loop is 
	terminated and the next iteration of the outer loop starts (lines \ref{line:algo-if-t-null}--\ref{line:algo-continue}).
	Otherwise, the new test case is added 
	to $T_{i,j}$ (line~\ref{line:algo-update-Tij}) and finally to $T_{i}$ (line~\ref{line:algo-update-Ti}). 
	Finally, the test-suite is reduced depending on parameter \textbf{RS} in line~\ref{line:algo-reduce-Ti}.

	\paragraph{Remarks.}
	For generating multiple \emph{different} test cases for the \emph{same} modification,
	only the set $T_{i,j}$ of already generated differentiating test cases for the current
	pair $P_i$, $P_j$ of program versions is taken into account.
	However, the approach may be generalized by blocking the whole set $T_i$ of
	already generated test cases in subsequent runs of the test-case generator. 
	This may, however, lead to drastically
	increased computational effort and is, therefore, 
	currently not supported.

	\subsection{Integration and Discussion of Strategies}
	The parameters \textbf{RS}, 
	\textbf{RTC}, \textbf{NRT}, \textbf{NPR} and \textbf{CR}
	allow for adjustments of efficiency and effectiveness 
	achievable by a regression-testing strategy concerning
	the detection of regression bugs.
	\paragraph{Efficiency.}
	Efficiency of regression-testing strategies may be measured
	in terms of (1) the computational effort (e.g., CPU time) 
	for selecting/generating regression test cases, 
	together with (2) the number of regression test cases being selected.
	Concerning (1), strategies with $\textbf{RTC}=\textbf{MT}$ are presumably more efficient
	than those with $\textbf{RTC}=\textbf{MR}$ as finding modification-revealing
	test cases is, on average, computationally more complicated than only reaching a modified program location. 
	Concerning (2), the number of regression test cases to be selected 
	for each new program version is (at most) $\textbf{NRT}\cdot\textbf{NPR}$
	thus presumably growing with increasing values of $\textbf{NRT}$ and $\textbf{NPR}$.
	Additionally, strategies with $\textbf{RS}\neq\textbf{None}$ are presumably 
	less efficient in terms of CPU time, but presumably much more efficient 
	in terms of the number of test cases.
	Finally, $\textbf{CR}$ also presumably decreases efficiency in terms 
	of CPU time, however presumably increases efficiency in terms of number of test cases.
	\paragraph{Effectiveness.}
	Effectiveness of regression-testing strategies may be measured
	in terms of the number of regression bugs detected by the selected regression test cases.
	Test cases selected for $\textbf{RTC}=\textbf{MR}$ are presumably
	more effective than those for $\textbf{RTC}=\textbf{MT}$.
	Similarly, fault-detection rates of regression test suites
	presumably increase with higher values for \textbf{NRT} (i.e., leading to
	more different ways of testing modifications between program versions)
	as well as \textbf{NPR} (i.e., taking more previous program versions
	into account).
	For instance, concerning the revision from $P_2$ to $P_3$ in our example,
	revealing bug~3 in $P_2$ requires a test case such as $t_2$.
	This can be ensured by strategies with $\textbf{RTC}=\textbf{MR}$,
	$\textbf{NRT}>0$, and $\textbf{NPR}>0$ which are, however, presumably less 
	efficient than strategies with $\textbf{RTC}=\textbf{MT}$.
	Finally, $\textbf{RS}$ and $\textbf{CR}$ might decrease effectiveness as 
	reducing the test suite might remove test cases that would succeed in finding the bug (see Sect.~\ref{sec:reduction-strategies}).

	In the next section, we empirically
	investigate these assumptions by presenting the results of an experimental 
	evaluation gained from applying a tool for our methodology
	to a collection of subject systems.

	\section{Experimental Evaluation}\label{sec5:eval}

	The configurable framework
	presented in the previous section allows us
	to investigate the impact of the 
	parameters $\protect\raisebox{.5pt}{\textcircled{\raisebox{-.5pt} {\small 1}}}$--$\protect\raisebox{.5pt}{\textcircled{\raisebox{-.5pt} {\small 5}}}$ (see Sect.~\ref{sec3:meth}) on efficiency
	and effectiveness of regression testing.
	In our experimental evaluation, we consider
	version histories of C program units (i.e., function level). 
	We use real-world systems obtained from GitHub and the change history 
	provided by git. 
	To systematically compare effectiveness of different regression-testing strategies,
	we further employ simulated bugs throughout program-version histories
	and measure the corresponding fault-detection ratio of the selected test suites.
	We do this by repeatedly applying standard mutation operators for C programs.
	Correspondingly, we measure efficiency in terms of the computational effort for generating regression test cases
	as well as in terms of the number of test cases to be (re-)executed throughout the version history.
	\subsection{Research Questions}
	We consider the following research questions.
	\begin{itemize}
		\item[\textbf{(RQ1)}] How does the regression-testing strategy impact \textit{testing effectiveness}?
		\item[\textbf{(RQ2)}] How does the regression-testing strategy impact \textit{testing efficiency}?
		\item[\textbf{(RQ3)}] Which regression-testing strategy constitutes the \textit{best trade-off between effectiveness and efficiency}?
	\end{itemize}
	\subsection{Experimental Setup}
	We next describe the evaluation methods and experimental design used in our
	experiments to address these research questions.
	\paragraph{Methods and Experimental Design}
	To compare different strategies for regression-testing, we instantiate the five parameters 
	$\textbf{RTC}$, $\textbf{NRT}$, $\textbf{NPR}$, $\textbf{RS}$ and $\textbf{CR}$
	of our methodology as described in Sect.~\ref{sec3:meth} as follows:
	\begin{itemize}
		\item $\textbf{RTC}\in\{\textbf{MT}, \textbf{MR}\}$ (RTS Generation Criterion).
		\item $\textbf{NRT}\in\{\textbf{1},\textbf{2},\textbf{3}\}$ (Number of Regression Test Cases per Revision).
		\item $\textbf{NPR}\in\{\textbf{1},\textbf{2},\textbf{3}\}$ (Number of Previous Program Revisions).
		\item $\textbf{RS}\in\{\textbf{None}, \textbf{ILP}, \textbf{FAST++}, \textbf{DIFF}\}$ (TSR Strategy).
		\item $\textbf{CR}\in\{\textbf{CR}, \textbf{No-CR}, \textbf{None}\}$ (Continuous Reduction).
	\end{itemize}
	To answer $\textbf{RQ1}$-$\textbf{RQ3}$, 
	each instantiation of all five parameters, therefore, corresponds to one particular \emph{regression-testing strategy}
	under consideration, where we denote a strategy by $[\textbf{RTC}, \textbf{NRT}, \textbf{NPR},\allowbreak \textbf{RS}, \textbf{CR}]$ for short.
	We thus obtain 144 reasonable regression-testing strategies (since the combination of $\textbf{RS}=\textbf{None}$ and $\textbf{CR}=\textbf{CR}$ is meaningless
	and we do not take into account strategies having $\textbf{CR}=\textbf{None}$ and $\textbf{RS}!=\textbf{None}$ 
	as reduction of non-accumulated test-suites is also meaningless).
	Two of these strategies may be considered as proper \emph{baselines}:
	\begin{itemize}
		\item Baseline 1 (basic regression testing strategy):$[\textbf{MT},\textbf{1},\textbf{1},\textbf{None},\textbf{No-CR}]$.
		\item Baseline 2 (basic regression testing without initial test suite):$[\textbf{MT},\textbf{1},\textbf{1},\textbf{None},\textbf{None}]$.
	\end{itemize}
	We limit our considerations to 3 as maximum 
	value for $\textbf{NRT}$ and $\textbf{NPR}$ as the increase of effectiveness 
	diminishes for higher values of $\textbf{NRT}$ and $\textbf{NPR}$. 
	Additionally, we do not consider $0$ as value for $\textbf{NRT}$ and $\textbf{NPR}$, 
	as this would obviously result in an empty test-suite.

	We further divide \textbf{RQ2} into the following sub-questions.
	\begin{itemize}
		\item \textbf{(RQ2.1)} How does the regression-testing strategy impact \textit{efficiency} in terms of CPU time?
		\item \textbf{(RQ2.2)} How does the regression-testing strategy impact \textit{efficiency} in terms of number of test-cases?
	\end{itemize}
	Note that we consider test-case generation time for CPU time and not test-case execution time. This is due to the fact, that test-case execution time was negligible during our evaluation.

	\paragraph{Subject Systems.}
	We consider program units as our subject systems in terms of \emph{testable functions} extracted from real-world C programs 
	for which we further require available version history.
	We therefore focus on open-source projects from GitHub. 
	We selected program units in terms of preprocessed \textit{c}-files 
	consisting of an entry-function (i.e., the function-under-test) as well as all 
	functions within the same compilation unit having (direct or indirect) callee-dependencies to 
	the function-under-test.
	To sum up, our subject systems have to fulfill the following requirements
	to be useable for our evaluation purposes.
	\begin{itemize}
		\item The unit must be processable by the test-case generator used in our framework (e.g., syntax must be ANSI C, no multi-threading, etc.).
		\item The functions-under-test (as well as all callees) 
		must have undergone a version-history of at least 3 modifications.
		\item The signature of the functions-under-test must provide 
		input parameters for which different values will yield different 
		return values (or affect the values of at least one global variable). 
		However, this needs not to be known in advance as the existence of such parameters is actually checked during test generation 
		(which may timeout if this requirement is not met).
		\item Calling the function-under-test multiple times with the same parameter values 
		produces the same return value or global-variable values 
		(i.e., no non-deterministic behavior or external system-function calls are considered).
	\end{itemize}
	
	The resulting collection of 
	subject systems comprises program units from open-source projects published in GitHub:
	\begin{itemize}
		\item \textbf{betaflight}\footnote{https://github.com/betaflight/betaflight} contains six program units from the flight controller software \emph{betaflight}. 
		\item \textbf{netdata}\footnote{https://github.com/netdata/netdata} contains six program units from the infrastructure monitoring and troubleshooting software \emph{netdata}.
		\item \textbf{wrk}\footnote{https://github.com/wg/wrk} contains one program unit from the HTTP benchmarking tool \emph{wrk}.
	\end{itemize}
	The size of subject systems ranges from 270 to 950 lines of code (after removing unnecessary code). 
	The number of changes (i.e., commits) for each subject system is between 4 and 18 (only counting commits changing code inside the unit).
	
	\paragraph{Simulating Bugs.} 
	
	Although software evolution and regression testing become more and more important, 
	properly documented and practically usable histories of program versions combined with
	information about real bugs in software projects are still barely available.
	Hence, for in-depth investigations of the interactions between program revisions, program bugs and their
	detection by regression test cases as required for answering our research
	questions, we have to rely on synthetically generated bugs.
	Mutation testing is a well-established
	approach for evaluating effectiveness of testing techniques by simulating
	common program faults~\cite{Andrews2005}. 
	In particular, mutation-testing tools provide
	collections of syntactic program-transformation operations such
	that the resulting modified program potentially shows different (i.e.,
	erroneous) output behaviors. 

\begin{figure}[tp]
		\centering
		\includegraphics[width=0.9\linewidth]{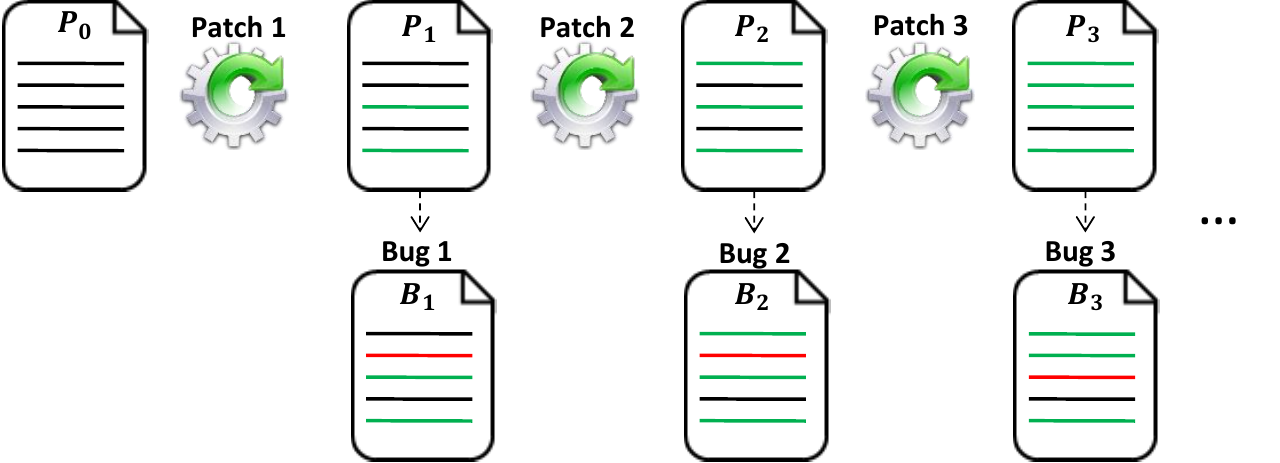}
		\caption{Bug Simulation using Mutation Testing}\label{fig:bugs}
	\end{figure}

	Fortunately, it has been recently shown that mutations provide
	a reliable substitute for realistic bugs in testing experiments~\cite{Andrews2005}.
	We, therefore, utilize concepts from mutation testing
	to simulate bugs in our subject systems.
	Figure~\ref{fig:bugs} provides an overview of our approach. 
	For each program version other than program version $P_0$, we create mutants containing simulated bugs.
	The regression-test generation is then executed on this bugged version and 
	effectiveness of the resulting test suite is measured by executing 
	the generated/selected regression-test suite on both the bugged and the bug-fixed (i.e., the original) version and by comparing the return 
	values (see Sect.~\ref{sec3:meth}).
	If the return values differ, the bug is successfully detected by the test-suite.
	We used 62 mutation operators in total (the names of the mutation operators are provided on our supplementary web page~\footnote{https://www.es.tu-darmstadt.de/regrets}), which can be clustered into three different types. 
	The first group consists of mutation operators replacing constants 
	and variables (e.g., replace variable $a$ by variable $b$, replace constant $5$ by constant $6$ etc.).
	The second group consists of mutation operators replacing operators (e.g., replacing $+$ by $-$).
	The last group consists of mutation operators replacing pointers and array references (e.g., replacing pointer $pt1$ by pointer $pt2$).
	
	\paragraph{Data Collection.}

	For comparing different regression-testing strategies, we first 
	generate a faulty revisions $B_i$ from all 
	revisions $P_i$ of each subject system by simulating bugs as described before.
	Next, we introduce program labels into each faulty revision $B_i$ 
	marking each line of code which has been modified since the last version $P_{i-1}$. 
	These labels are used as test goals to generate 
	modification-traversing test cases (similarly for $P_{i-2}$ and $P_{i-3}$ depending on parameter $\textbf{NPR}$).
	In addition, we generate the \textit{comparator}-programs described 
	in Sect.~\ref{sec3:meth} for generating modification-revealing test-cases. 
	The \textit{comparator}-program is further used for comparing test-case executions on the faulty version $B_i$ 
	and the previous versions $P_{i-1}$, $P_{i-2}$ and $P_{i-3}$. 
	If no \textit{comparator}-program can be created (e.g., due to the limitations explained in Sect.~\ref{sec3:meth}), 
	no test suite will be generated for all strategies targeting this specific comparator-program.

	We applied Algorithm~\ref{algorithm:tcgen}
	to generate regression-test suites $T_i^{s}$ for every bugged revision $B_i$ 
	using all 144 strategies.

	To answer \textbf{RQ1}, we measure effectiveness of the generated test suites such that
	test suite $T_i^{s}$ detects the bug in $B_i$ if
	it contains at least one test case $t$ whose execution yields
	different output values when executed on $P_i$ and $B_i$
	(denoted $P_i(tc) \neq B_i(tc)$).
	Hence, effectiveness with respect to a test suite $T_i^{s}$
	is calculated as 
	\setlength{\abovedisplayskip}{1.5pt}
	\setlength{\belowdisplayskip}{1.5pt}
	\setlength{\abovedisplayshortskip}{1.5pt}
	\setlength{\belowdisplayshortskip}{1.5pt}
	\begin{equation}
		\textit{effectiveness}(T_i^{s})= \textit{detects}(T_i^{s}, i)
		\label{effectivenessTestSuite}
	\end{equation}
	where $\textit{detects}(T_i^{s}, i) = 1 $ if $\exists t \in T_i^{s}:P_i(t) \neq B_i(t)$, or 0 otherwise, and $B_i$ is the faulty revisions generated for $P_i$.
	Effectiveness of a strategy \textit{s} is calculated as
	\begin{equation}
		\textit{effectiveness}(s)= \frac{\sum_{i=0}^{n}  \textit{effectiveness}(T_i^{s})}{n}
		\label{effectivenessStrategy}
	\end{equation}
	where $n$ is the number of revisions.

	To answer $\textbf{RQ2}$, we measure efficiency in two ways. 
	First, efficiency of regression 
	test suites generated by strategy $s$ for program unit $P_i$ is calculated
	as the average \emph{size} of test suites $T_i^s$ in terms of the number
	of test cases
	\begin{equation}
		\textit{efficiency}_{\textit{size}}(T_i^{s})=|T_i^{s}|
		\label{efficiencyTSSize}
	\end{equation}
	such that the overall efficiency of strategy \textit{s} is calculated as
	\begin{equation}
		\textit{efficiency}_{\textit{size}}(s)=\frac{\sum_{i=0}^{n}  \textit{efficiency}_\textit{size}(T_i^{s})}{n}.
		\label{efficiencyStrategySize}
	\end{equation}
	Second, efficiency of the regression test-suite generation is given 
	as the aggregated CPU time $\textit{efficiency}_{\textit{CPU}}$
	consisting of the initialization phase of the software model-checker, the
	reachability analysis, and the test-case extraction.
	While the first phase is only executed once for each program revision 
	under test, the other phases are executed by the number of test goals multiplied by parameter $\textbf{NRT}$.
	Additionally, for practical reasons, the whole process is limited by a global time-out parameter after which 
	the process terminates without providing further test cases thus potentially 
	leading to less test cases than specified by parameter $\textbf{NRT}$.
	The equation for calculating $\textit{efficiency}_{\textit{CPU}}$ of the different strategies is 
	\begin{equation}
		\textit{efficiency}_{\textit{cpu}}(s)=\frac{\sum_{i=0}^{n}  \textit{efficiency}_\textit{cpu}(T_i^{s})}{n}.
		\label{efficiencyCPUStrategySize}
	\end{equation}
	Finally, to answer $\textbf{RQ3}$, we calculate the mean values of each strategy 
	for each program unit in our subject system for effectiveness and efficiency and, 
	thereupon, calculate the trade-off as
	\begin{equation}
		\textit{trade-off}(s)=\frac{\textit{effectiveness}(s)}{\textit{efficiency}_\textit{size}(s)}.
	\end{equation}

	\paragraph{Tool Support.}
	We implemented Algorithm~\ref{algorithm:tcgen} in a tool, 
	called \textsc{RegreTS} (\emph{Regression Testing Strategies}).
	\textsc{RegreTS} extends the software 
	model-checker \textsc{CPAchecker} for C programs to 
	generate regression test cases. 
	This is achieved by performing reachability-analysis runs 
	for program locations marked as test goals (see Sect.~\ref{sec3:meth}).
	Additionally, we use \textsc{MUSIC}~\cite{music}, a mutation
	tool for C programs, to generate synthetic bugs as described above.
	For measuring test coverage, we utilize 
	\textsc{TestCov}\footnote{https://gitlab.com/sosy-lab/software/test-suite-validator}, 
	a test-suite executor for C programs.
	\textsc{TestCov} further supports TSR with a greedy algorithm, as explained 
	in Sect.~\ref{sec3:meth}, called $\textbf{DIFF}$.
	Furthermore, we extended \textsc{TestCov} to also support $\textbf{FAST++}$ and $\textbf{ILP}$.
	All tools are available on our website\footnote{https://www.es.tu-darmstadt.de/regrets}.

	\paragraph{Measurement Setup.}
	\textsc{RegreTS} as well as \textsc{MUSIC} and 
	\textsc{TestCov} have been executed on an Ubuntu 18.04 machine, equipped 
	with an Intel Core i7-7700k CPU and 64~GB of RAM.
	The CPU time for test-suite generation 
	is limited to 900~s per revision, and the CPU time for the execution of the test cases to detect 
	bugs is limited to 30~s per test case.
	The TSR is not limited by CPU time as the CPU time needed for reduction was negligible.
	We executed our evaluation with Java 1.8.0-171 and 
	limited the Java heap to 15~GB.
	

	\subsection{Results}
\begin{figure}
		\centering
		\subfloat[Strategies with $\textbf{RTC}=\textbf{MR}$]{\label{fig:results-mr}\includegraphics[angle=270,origin=c, width = 0.44\textwidth]{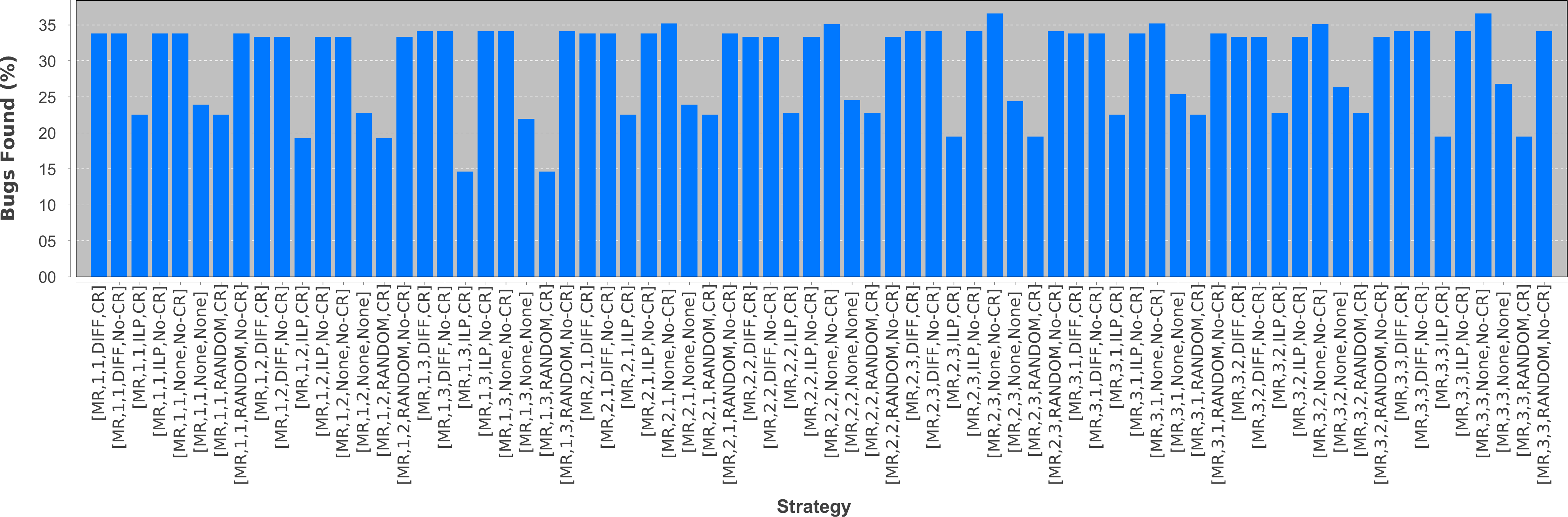}}
		\hfill
		\subfloat[Strategies with $\textbf{RTC}=\textbf{MT}$]{\label{fig:results-mt}\includegraphics[angle=270,origin=c,width = 0.44\textwidth]{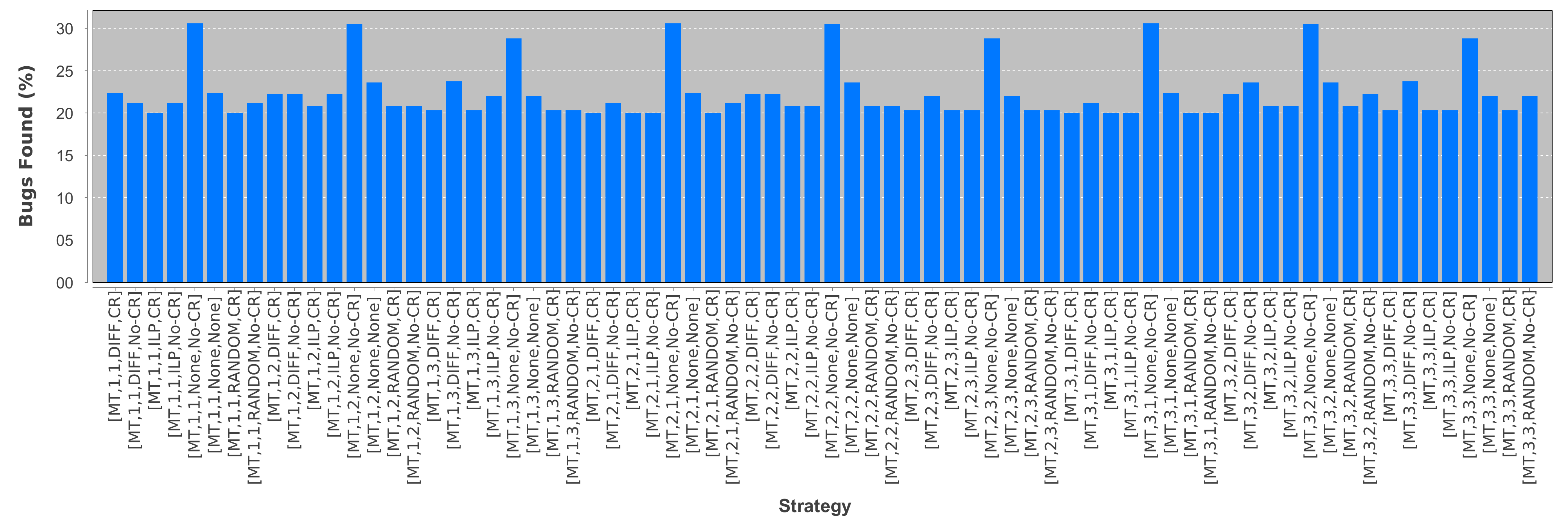}}
		\caption{Results Effectiveness}\label{fig:results}
	\end{figure}

\begin{figure}
		\centering
		\subfloat[Strategies with $\textbf{RTC}=\textbf{MR}$]{\label{fig:results-mr-gentime}\includegraphics[angle=270,origin=c, width = 0.44\textwidth]{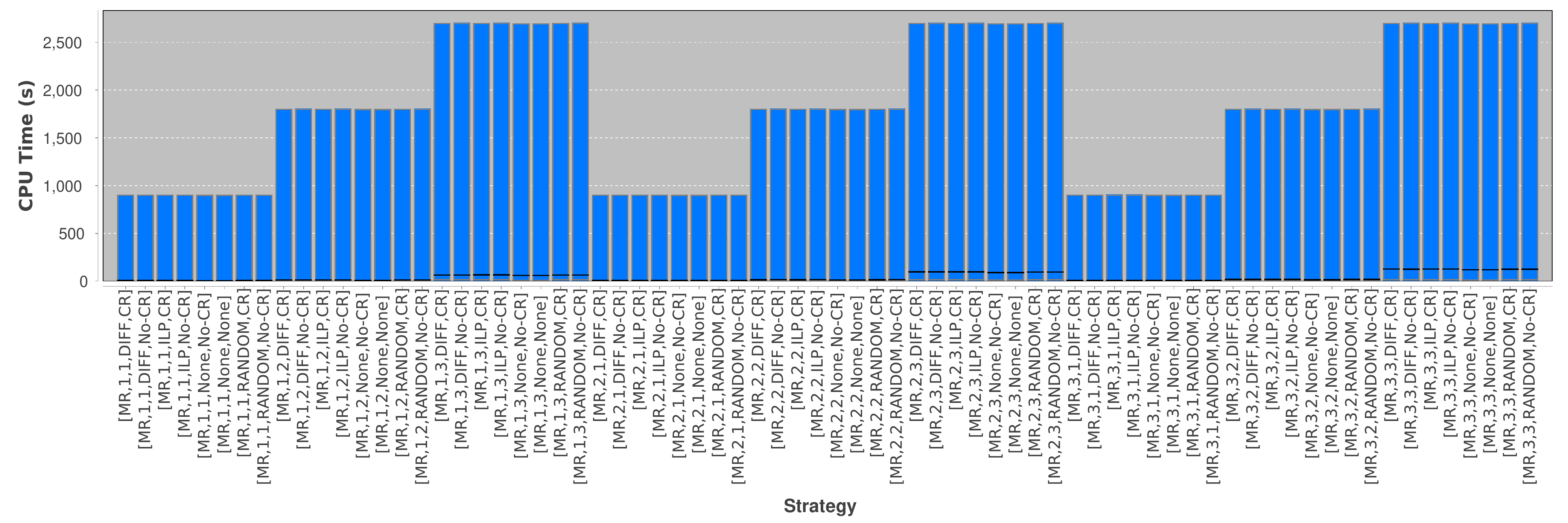}}
		\hfill
		\subfloat[Strategies with $\textbf{RTC}=\textbf{MT}$]{\label{fig:results-mt-gentime}\includegraphics[angle=270,origin=c, width = 0.44\textwidth]{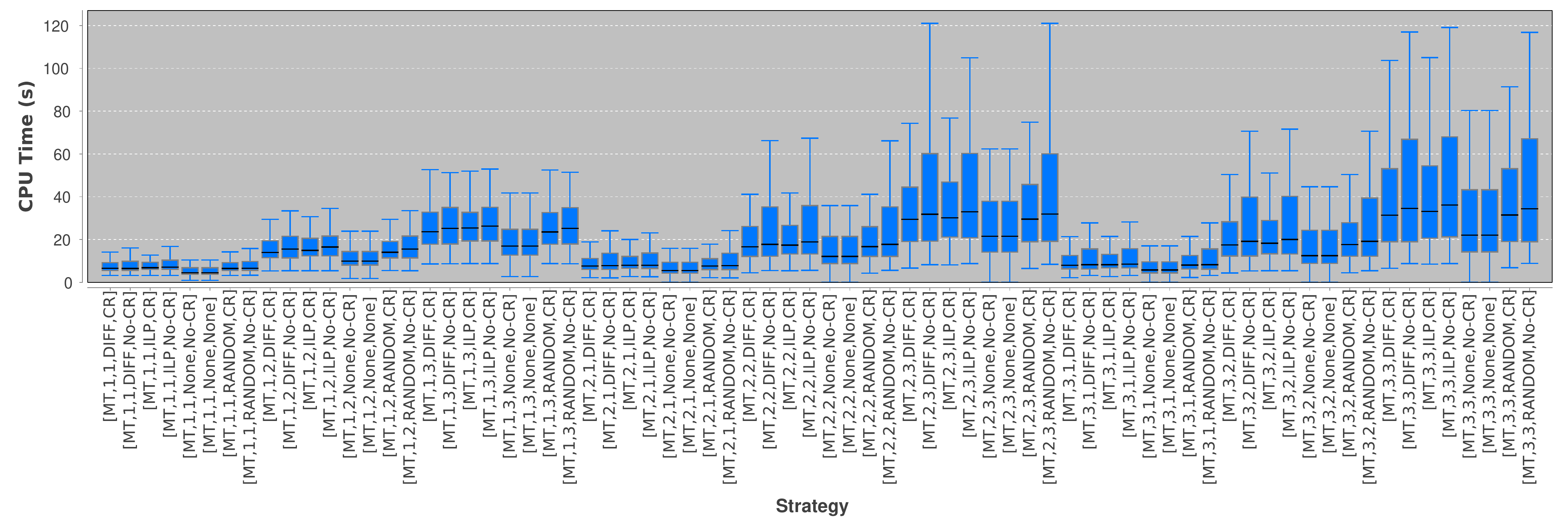}}
		\caption{Results Generation Time}\label{fig:results_gentime}
	\end{figure}

\begin{figure}
		\centering
		\subfloat[Strategies with $\textbf{RTC}=\textbf{MR}$]{\label{fig:results-mr-size}\includegraphics[angle=270,origin=c, width = 0.44\textwidth]{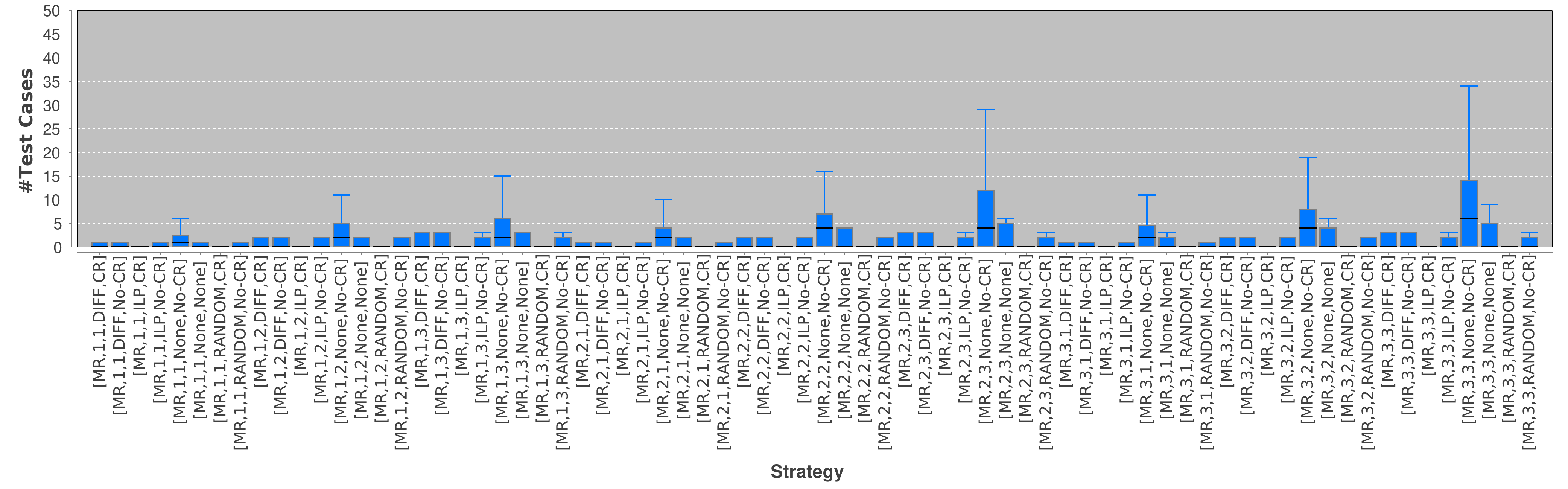}}
		\hfill
		\subfloat[Strategies with $\textbf{RTC}=\textbf{MT}$]{\label{fig:results-mt-size}\includegraphics[angle=270,origin=c, width = 0.44\textwidth]{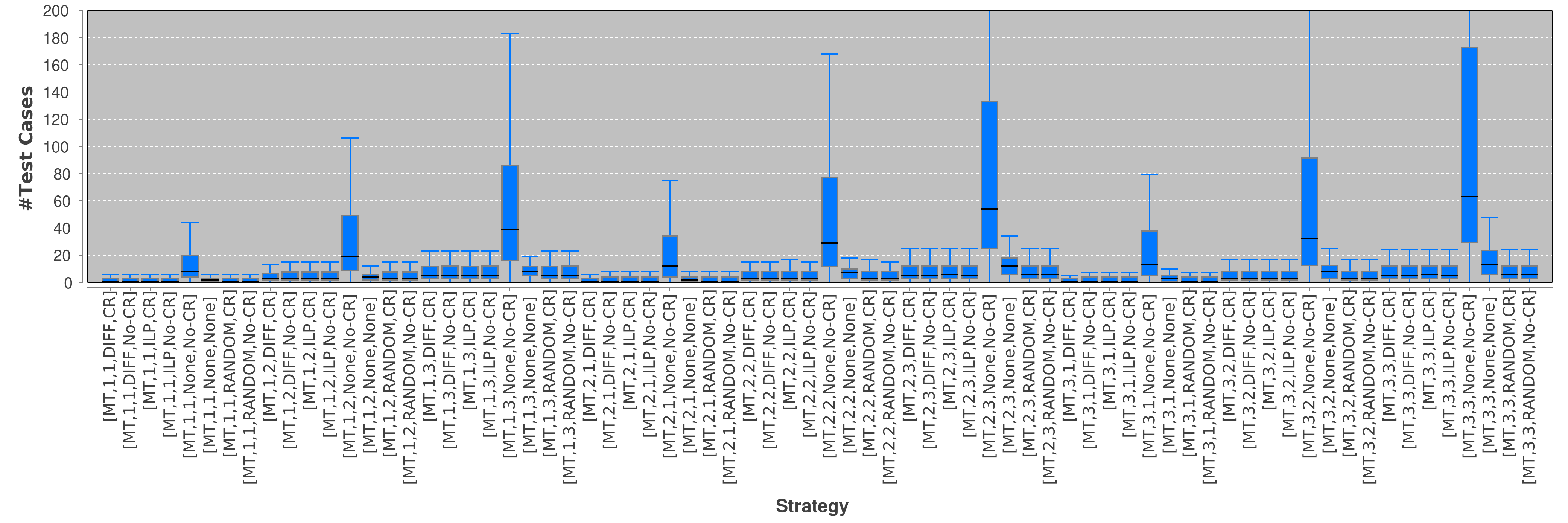}}
		\caption{Results Test-Suite Size}\label{fig:results_size}
	\end{figure}

\begin{figure}
		\centering
		\subfloat[Strategies with $\textbf{RTC}=\textbf{MR}$]{\label{fig:results-mr-tradeoff}\includegraphics[angle=270,origin=c, width = 0.44\textwidth]{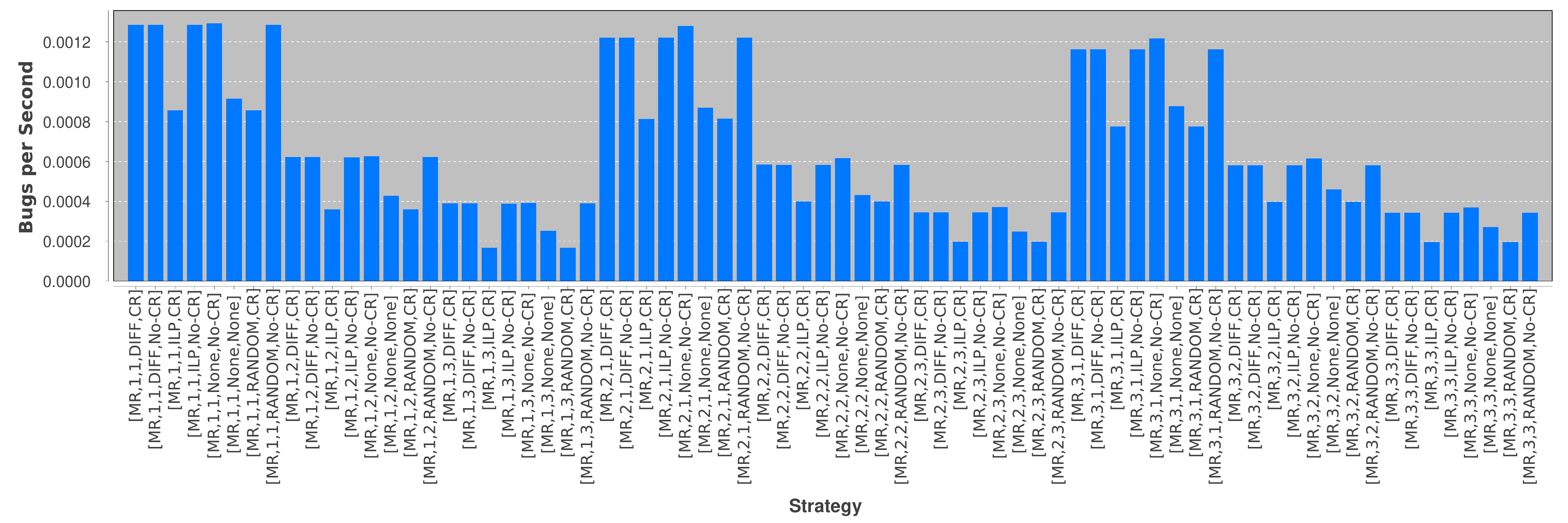}}
		\hfill
		\subfloat[Strategies with $\textbf{RTC}=\textbf{MT}$]{\label{fig:results-mt-tradeoff}\includegraphics[angle=270,origin=c, width = 0.44\textwidth]{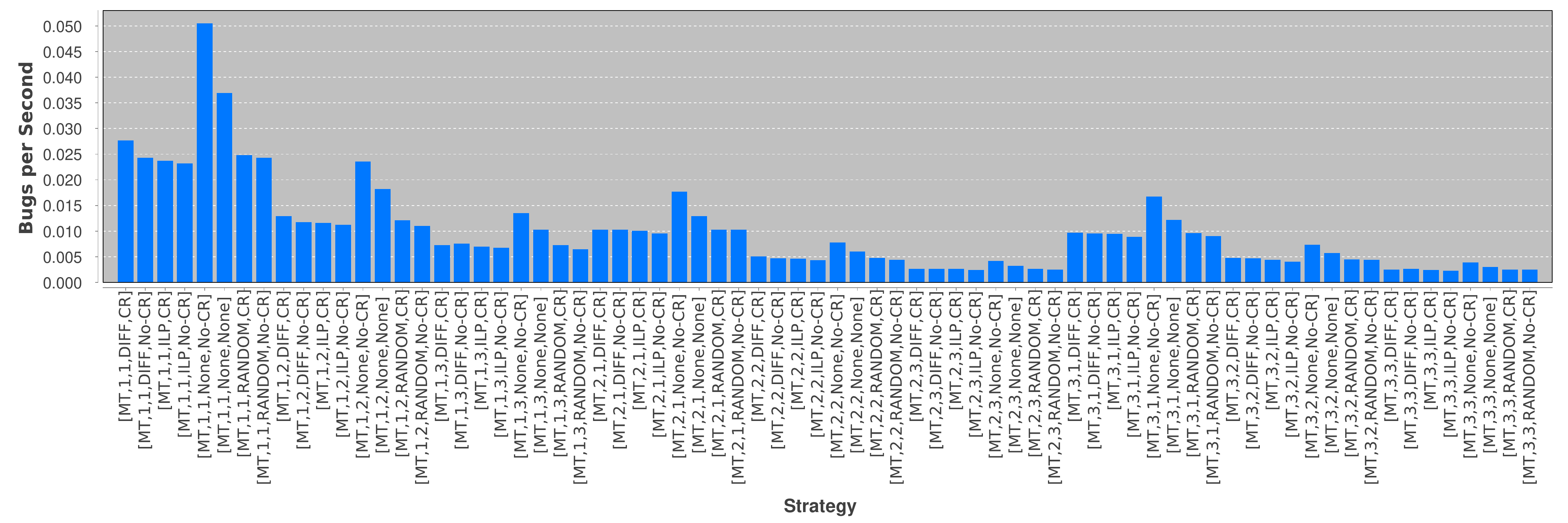}}
		\caption{Results Bugs per Second}\label{fig:results_tradeoff}
	\end{figure}

	The measurements for \textbf{RQ1} are depicted in Fig.~\ref{fig:results-mr} 
	for all strategies with $\textbf{RTC} = \textbf{MR}$ and in Fig.~\ref{fig:results-mt} for all strategies with $\textbf{RTC} = \textbf{MT}$.
	Results for \textbf{RQ2} are shown in Figs.~\ref{fig:results-mr-gentime} and \ref{fig:results-mr-size} for all strategies with $\textbf{RTC}=\textbf{MR}$ and Figs.~\ref{fig:results-mt-gentime} and \ref{fig:results-mt-size} for all strategies with $\textbf{RTC}=\textbf{MT}$.
	The results for \textbf{RQ3} are shown in Fig.~\ref{fig:results-mr-tradeoff} for all strategies with $\textbf{RTC}=\textbf{MR}$ and Fig.~\ref{fig:results-mt-tradeoff} for all strategies with $\textbf{RTC}=\textbf{MT}$.
	The box plots in Figs.~\ref{fig:results-mr-gentime}, \ref{fig:results-mt-gentime}, \ref{fig:results-mr-size} and \ref{fig:results-mt-size}  aggregate the results after applying the formulas~\ref{efficiencyStrategySize} and \ref{efficiencyCPUStrategySize} (see above) for all subject systems. 
	The boxes depict the range of results while the black dashes depict the mean values. 
	The whiskers show the minimum and maximum values (excluding outliers).
	
	\paragraph{RQ1 (Effectiveness).}
	The best effectiveness for our subject
	systems is achieved by the strategies $[\textbf{MR}, \textbf{3}, \textbf{3},\textbf{None}, \allowbreak \textbf{No-CR}]$, 
	$[\textbf{MR}, \textbf{2}, \textbf{3},\textbf{None},\allowbreak \textbf{No-CR}]$, 
	$[\textbf{MR}, \textbf{2}, \textbf{3},\textbf{None},\\ \textbf{No-CR}]$, 
	$[\textbf{MR}, \textbf{3}, \textbf{3}, \textbf{FAST++},\allowbreak \textbf{CR}]$ and
	$[\textbf{MR}, \textbf{3}, \textbf{3},\textbf{FAST++}, \textbf{No-CR}]$
	 with an average bug detection rate of 0.3659. 
	Compared to the baseline 1
	$[\textbf{MT},\textbf{1},\textbf{1},\textbf{None},\allowbreak \textbf{No-CR}]$ with an average detection rate of 0.306
	and baseline 2
	$[\textbf{MT},\textbf{1},\textbf{1},\textbf{None},\textbf{None}]$  with an average detection rate of 0.224. 
	The worst performing strategy is $[\textbf{MT}, \textbf{3}, \textbf{3},\textbf{FAST++}, \textbf{CR}]$ with an average detection rate of 0.119.
	
	\paragraph{RQ2 (Efficiency).} 
	The best efficiency measure concerning CPU
	time is obtained by strategy $[\textbf{MT},\textbf{1},\textbf{1},\textbf{None},\textbf{No-CR}]$ and $[\textbf{MT},\textbf{1},\textbf{1},\textbf{None},\textbf{None}]$ 
	with an average amount of 6.058s. 
	The worst performing strategy, $[\textbf{MR},\textbf{3},\textbf{3},\textbf{ILP},\textbf{No-CR}]$, requires 993.229s on average. 
	Concerning test-suite sizes, $[\textbf{MT},\textbf{1},\textbf{1},\textbf{ILP},\textbf{CR}]$, $[\textbf{MT},\textbf{2},\textbf{1},\textbf{ILP},\textbf{CR}]$, and $[\textbf{MT},\textbf{3},\textbf{1},\textbf{ILP},\textbf{CR}]$
	perform best with an average measure of 0.225 test cases, whereas
	strategy $[\textbf{MT},\textbf{3},\textbf{3},\textbf{None},\allowbreak \textbf{No-CR}]$ leads to the largest test-suite sizes with an
	average measure of 115.44 test cases.
	
	\paragraph{RQ3 (Trade-Off).}
	The base trade-off between effectiveness 
	and CPU time is obtained by strategy $[\textbf{MT},\textbf{1},\textbf{1},\textbf{None},\textbf{No-CR}]$ with an average of 0.0503 bugs found per second.
	The worst performing strategy is $[\textbf{MT},\textbf{3},\textbf{3},\textbf{FAST++},\textbf{CR}]$ with an average of 0.0015 bugs found per second.
	

	\subsection{Discussion and Summary}
	
	\paragraph{RQ1 (Effectiveness).}
	
	Setting parameter $\textbf{RTC}$ to $\textbf{MR}$ increases effectiveness for all strategies. 
	In fact, almost all strategies using $\textbf{MR}$ are more effective than even 
	the most effective strategy using $\textbf{MT}$.
	Indeed, parameter $\textbf{RTC}$ has the highest impact on effectiveness.
	While the impact of other parameters is smaller, it is nonetheless also observable.
	Choosing $\textbf{None}$ for parameter $\textbf{RS}$ increases effectiveness, as the other 
	parameter value reduces effectiveness by reducing the number of test cases.
	Choosing $\textbf{CR}$ or $\textbf{No-CR}$ for parameter $\textbf{CR}$ 
	has nearly no impact on effectiveness. 
	Strategies with $\textbf{MR}$ and $\textbf{ILP}$ cause a significant loss in effectiveness 
	if $\textbf{CR}$ is selected.
	Choosing $\textbf{None}$ for parameter $\textbf{CR}$ has a negative impact on strategies with $\textbf{MR}$ selected, 
	however, only having a small impact if $\textbf{MT}$ is selected.
	Lastly, parameters $\textbf{NRT}$ and $\textbf{NPR}$ also increase effectiveness with increasing values (even in case of small increases).

	\noindent%
	\begin{tabularx}{\textwidth}{|X|}%
		\hline
		\cellcolor{black}\textcolor{white}{Answer \textbf{RQ1}}\\
		\hline
		\cellcolor{green!20}The best effectiveness measure is reached by strategy $[\textbf{MR}, \textbf{3}, \textbf{3},\textbf{None}, \textbf{No-CR}]$ (i.e., modification revealing test cases, three test cases per test goal, up to three previous revisions and no test-suite reduction) which improves effectiveness compared to the baseline $[\textbf{MT},\textbf{1},\textbf{1},\textbf{None},\textbf{No-CR}]$ by ~19\% and compared to the baseline $[\textbf{MT},\textbf{1},\textbf{1},\textbf{None},\textbf{None}]$ by ~61\%.
		The highest impact on effectiveness is caused by parameter $\textbf{RTC}$\\
		\hline
	\end{tabularx}%
	\noindent%
	
	\paragraph{RQ2.1 (CPU Time).}
	Parameter $\textbf{RTC}$ has by far the highest impact on CPU time. 
	When choosing $\textbf{MT}$, CPU time increases nearly 20 fold.
	As expected, parameter $\textbf{NPR}$ increases the CPU time nearly linearly to its respective value.
	Unexpectedly, parameter $\textbf{NRT}$ does not affect CPU time by a large margin.
	This is most likely due to the fact, that reachability information 
	computed during the first run of the test generator can be re-used for the next test cases.
	Parameters $\textbf{CR}$ and $\textbf{RS}$ are negligible in terms of 
	CPU time if $\textbf{MR}$ is selected, since the number of test-cases 
	remains small. 
	If $\textbf{MT}$ is selected, both parameters impact CPU time, 
	however still only by a small margin.

	\noindent%
	\begin{tabularx}{\textwidth}{|X|}%
		\hline
		\cellcolor{black}\textcolor{white}{Answer \textbf{RQ2.1}}\\
		\hline
		\cellcolor{green!20}The best efficiency measure in terms of CPU time is reached by the baseline 
		strategies $[\textbf{MT},\textbf{1},\textbf{1},\textbf{None},\textbf{No-CR}]$ and $[\textbf{MT},\textbf{1},\textbf{1},\textbf{None},\textbf{None}]$ (i.e., modification traversing test cases, one test case per test goal, one previous revision and no test-suite reduction, either ignoring or using the previous test-suite (as this makes no different in CPU time)).
		The highest impact on CPU time is caused by parameter $\textbf{RTC}$.\\
		\hline
	\end{tabularx}%
	\noindent%

	\paragraph{RQ2.2 (Test-Suite Size).}
	The highest impact on efficiency in terms of test-suite size is caused by 
	parameter $\textbf{RS}$. 
	If $\textbf{None}$ is selected, no reduction is enabled, and therefore, 
	the test-suite grows with each version.
	However, the choice of the technique used for test-suite 
	reduction only has a small impact on the test-suite size.

	Parameter $\textbf{RTC}$ also has a high impact on the test-suite size.
	This is due to the fact, that a patch might contain multiple modifications. 
	Therefore, the number of test cases is higher in case of $\textbf{MT}$
	(i.e., requiring one test-case per modified line) as compared to $\textbf{MR}$, where only one test case is required.
	Parameter $\textbf{CR}$ also affects test-suite sizes, but only by a small amount.
	Lastly, parameters $\textbf{NRT}$ and $\textbf{NPR}$ affect the test-suite size 
	almost linearly to their respective values.
	
	\noindent%
	\begin{tabularx}{\textwidth}{|X|}%
		\hline
		\cellcolor{black}\textcolor{white}{Answer \textbf{RQ2.2}}\\
		\hline
		\cellcolor{green!20}The best efficiency measure in terms of test-suite size is reached by the strategy $[\textbf{MR},\textbf{1},\textbf{1},\textbf{ILP},\textbf{CR}]$ (i.e., modification revealing test cases, one test case per test goal, one previous revision, ILP as test-suite reduction strategy and using the reduced test-suite of the previous revision)
		which improves effectiveness compared to the baseline $[\textbf{MT},\textbf{1},\textbf{1},\textbf{None},\textbf{No-CR}]$ by ~6200\% and compared to the baseline $[\textbf{MT},\textbf{1},\textbf{1},\textbf{None},\textbf{None}]$ by ~970\%.
		The highest impact on the test-suite size is caused by parameter $\textbf{RS}$.\\
		\hline
	\end{tabularx}%
	\noindent%

	\paragraph{RQ3 (Trade-Off).}
	We observe that all parameters have a large impact on the trade-off between effectiveness and CPU time.
	Therefore, the interactions between the different parameters is the main driver 
	affecting the trade-off.
	
	\noindent%
	\begin{tabularx}{\textwidth}{|X|}%
		\hline
		\cellcolor{black}\textcolor{white}{Answer \textbf{RQ3}}\\
		\hline
		\cellcolor{green!20}The strategy yielding the best trade-off is $[\textbf{MT},\textbf{1},\textbf{1},\textbf{None},\textbf{No-CR}]$ (i.e., modification traversing test cases, one test case per test goal, one previous revisions, no test-suite reduction and using the test-suite of the previous revision as well)
		for which the fault-detection capability is acceptable, but the efficiency in terms of CPU time is very high.
		Compared to the second baseline $[\textbf{MT},\textbf{1},\textbf{1},\textbf{None},\textbf{None}]$ the trade-off is increased by 
		~36\%.\\
		\hline
	\end{tabularx}%
	\noindent%
	\paragraph{Remarks.}
	Strategy $[\textbf{MT}, \textbf{3},\textbf{3},\allowbreak \textbf{None},\allowbreak \textbf{No-CR}]$ is less effective than strategy $[\textbf{MT}, \textbf{3},\textbf{2},\\ \textbf{None},\allowbreak \textbf{No-CR}]$ 
	which, by definition, is counter-intuitive. 
	This is due to the fact, that some \textit{comparator}-programs 
	for comparing a faulty version $B_i$ to a program revision $P_{i-2}$ are invalid (e.g., due to different return types), whereas 
	the \textit{comparator}-program for $B_i$ and program revisions $P_{i-3}$ are actually valid. 
	However, no test cases could be found by the test-case generator revealing the bug.
	Hence, the number of test suites for $[\textbf{MT}, \textbf{3},\textbf{3}, \textbf{None},\allowbreak \textbf{No-CR}]$ and $[\textbf{MT}, \textbf{3},\textbf{2}, \textbf{None},\textbf{No-CR}]$ differ slightly, 
	which is the reason for the discrepancy between the factual results and the theoretical specification of the technique.
	In our subject systems, the probability 
	that no valid \textit{comparator}-program can be generated is approximately 8\%.
	Therefore, while this technical limitation is present, it should not affect the results by a large margin.
	
	\subsection{Threats to Validity}
	\paragraph{Internal Validity.}
	Our regression-testing methodology relies on the assumption that 
	when different program versions are tested with the same test case, all factors (e.g., platform, environment) that might 
	influence the output except for the source code itself remain constant.
	This so-called \emph{controlled-regression-testing assumption}
	is commonly used in regression-testing experiments and does, therefore,
	not harm validity of the results~\cite{Yoo2012}.

	Concerning the soundness of our methodology,
	we tested the test-generation loop by manually
	checking results for selected subject systems.
	However, due to undecidability of reachability of program locations,
	if no more test cases can be found (e.g., due to time-outs or imprecise counter-examples), 
	it is unknown if further test cases exist.
	Nevertheless, we expect precision improvements to not
	substantially obstruct the (relative) results of our evaluation.

	Another threat to validity might arise from our selection of mutation operators
	and their applications to our subject systems.
	However, our selection comprises those mutations leading to useful results 
	w.r.t. our experimental setting, namely
	affecting one line of code, performing no code deletions and 
	producing a compilable result.

	Limiting our considerations to (functional) unit testing may also threaten
	internal validity.
	As unit testing constitutes the most established and relevant testing
	technique in practice, it is particularly interesting and relevant to investigate our
	methodology at this level first.
	In addition, the proposed concepts might be likewise applicable at integration-
	and system-level.

	Additionally, our methodology does not incorporate systematic reusability-checks 
	of existing test cases for revealing modifications also in later revisions which may
	affect efficiency measures.
	We plan to extend our approach, accordingly, in a future work but we expect
	similar results as in our current setting.

	Finally, we expect our current focus on C programs to also not seriously
	harm validity as we expect similar results for other programming languages, at least for
	those relying on an imperative core (e.g., most OO languages).

	\paragraph{External Validity.}
	We are not aware of any competitive tools with similar functionality as \textsc{RegreTS},
	especially concerning the generation of a configurable number of modification-revealing
	test cases.
	Surprisingly, it was not possible for us to use other recent test-case generators 
	for strategies with $\textbf{RTC}=\textbf{MR}$, which, by design, should have been possible. 
	This might be due to the fact, that the subject systems are real-world programs, which where explicitly selected 
	to be processable by our test-case generator. 
	However, test-case generators are (usually) limited in supporting certain constructs of the C language. 
	We tried to use other test-case generators (i.e., \textsc{Klee}, \textsc{FuSEBMC}, \textsc{Symbiotic} and \textsc{PRTest}) 
	from the international testing competition 2021~\cite{Testcomp21}. 
	However, these test-case generators were barely able to generate any test cases at all. 
	In addition, successfully generated test cases were actually unable to reveal modifications between 
	different program version and were thus immediately removed from the test-suites during 
	TSR (leading to empty test suites). 
	Only \textsc{PRTest} was able to generate some meaningful test-cases but was, however, also
	not able to generate any test case for more than half of the subject systems thus being unusable for a proper comparison. 
	However, the main focus of this paper is to compare different regression-testing strategies, 
	and not to compare different test-case generators for regression-test generation. 
	We thus assume the results to be very similar 
	for other test-case generation techniques (even though effectiveness in terms of 
	CPU time might change, the ratio of the CPU time of different strategy presumably stay similar).

	Another threat might arise from the selection of subject systems
	and the usage of simulated bugs.
	Unfortunately, real-world systems with sufficient information about revisions and bugs as required 
	for our experiments are barely available.
	We evaluated three prominent candidates for potential candidates.
	First, \textsc{CoreBench} only provides a very short version history
	(often only 1--2 versions) and incorporates many bugs being 
	undetectable at unit level (e.g., involving files and
	global errors like overflows)~\cite{corebench}.
	Second, the \emph{regression-verification tasks} from the SV-Benchmarks~\cite{svbenchmarks} also have a 
	small version history and the different tasks cannot be executed 
	in a self-sustained manner as needed to reveal those bugs.
	However, we spend a lot of time in searching for other 
	freely available community benchmarks including version history and known bugs.
	However, suitable benchmarks are still very rare.
	Amongst others, we had a look into further subject systems from the SIR Repository including
	programs like gzip and make \cite{sir}, but either our test-case generator was not able
	to handle those programs for mostly technical reasons, or the programs were not suitable for our purposes (see descriptions above).
	As also already discussed above, mutation testing is a reliable fault-injection technique for measuring
	effectiveness in testing experiments~\cite{Andrews2005}.

	Finally, our tool relies on third-party software, namely \textsc{CPAchecker}, 
	a software model-checker, and the mutation tool MUSIC for C programs.
	However, both tools are established and have been extensively used
	for other experiments in the recent past, so we expect them to 
	produce sound results.

	\section{Related Work}\label{sec6:rw}
	\subsection{Regression-Testing}
	A comprehensive overview about regression testing
	is provided by \cite{Yoo2012}, describing 
	three categories:
	(1) \emph{minimization} of test suites as well as (2) 
	\emph{selection} and (3) \emph{prioritization} of test cases for regression testing.

	\paragraph{Test-suite minimization} is concerned with selecting from an existing test suite 
	a subset of test cases to reduce the number of \emph{redundant} test executions during regression testing.
	Many works propose heuristics for approximating near-optimal solutions~\cite{Chen1996,Harrold1993,Horgan1992,Offutt1995} 
	for this NP-complete optimization problem, requiring as inputs an \emph{existing} test suite and \emph{a-priori defined}
	metrics for measuring effectiveness of test cases.
	The approaches used in this paper for TSR have been proposed before.
	The greedy algorithm as used by \emph{DIFF} has been introduced in~\cite{TestCov}. To use ILP solving for TSR was initially proposed by~\cite{Khalilian2012} and FAST++ has been introduced by \cite{Cruciani2019}.
	However, none of these works investigate the interactions between RTS and TSR techniques as done in this paper.

	\cite{Shi_2018} evaluate existing test-suite-reduction techniques on 
	real-word projects based on their failed builds.
	\cite{Chen_2017} proposed a new technique to reduce test suites based on 
	assertions instead of structural code coverage 
	to improve effectiveness of the resulting test suite.
	In an earlier work, \cite{Shi_2015} compare and combine TSR and test-case 
	selection to further increase efficiency of regression testing.
	Our methodology goes beyond their approach as we consider further strategic parameters which turned
	out to be very relevant. 

	\paragraph{Regression-test selection} is concerned with selecting from an existing test suite of an evolving 
	program a subset of test cases to be (re-)executed on a new program version. 
	A variety of different techniques
	has been applied (e.g., control-flow analysis~\cite{Hartmann1989, Hartmann1990, Hartmann1990Revalidation}
	and/or data-flow analysis~\cite{Gupta1992,Harrold1988,Harrold1989Interprocedual,Taha1989}).
	Other works take behavior-preserving modifications (e.g., refactorings) 
	into account~\cite{Wang2018}, apply RTS to highly-configurable software~\cite{Marijan2018}, 
	and try to find pareto-optimal solutions for multi-objective RTS~\cite{Choudhary2018}.
	However, none of these works aim at \emph{generating new modification-revealing test cases} 
	to enhance effectiveness of regression testing as done in our work.
	In particular, most recent works only guarantee test cases to be modification-traversing.

	\paragraph{Test-case prioritization} is concerned with selecting from an existing test suite 
	a \mbox{(re-)}test-execution order such that effectiveness of testing 
	increases as quickly as possible over time (e.g., to find as many faults as fast as possible)~\cite{Wong1995}.
	The underlying problem is very similar
	to minimization/selection problems. 
	Most existing approaches consider code coverage as effectiveness criterion to statically 
	compute an (a-priori) ordering among test cases~\cite{Rothermel1999,Rothermel2001,Elbaum2001,Elbaum2001Costs,Malishevsky2002,Rothermel2002}.
	In a recent work, Wang and Zeng propose 
	a dynamic prioritization technique based on fault-detection history and 
	other properties~\cite{Wang2016}.
	In contrast, in our methodology, prioritization is currently out of scope, but may be easily
	incorporated during test-case generation using recent approaches.
	
	%
	%
	
	\subsection{Test-Case Generation}
	
	A wide range of technique exist for (automatically)
	generating test cases which we will describe in the following grouped by
	the test-case generation technique applied.
	However, we are not aware of related works in terms of the multiple test-cases 
	per test goal to increase effectiveness of the resulting test suite.
	
	\subsubsection{Coverage-Based Test-Case Generation}
	
	\paragraph{Fuzzing.} Fuzzing is currently very popular both in research and practice. 
	The idea is to quickly generate a large number of test cases by generating (semi-)random input values. 
	In some approaches, the input values of existing test cases are reused and modified 
	to generate new test cases.~\cite{FuzzingOverview}
	Recent fuzzing techniques are based on evolutionary algorithms~\cite{evolutionaryFuzzing} 
	or context-free grammars of the input data~\cite{grammarbasedFuzzing}.
	In addition, grey-box fuzzing~\cite{aflFuzzer} and whitebox fuzzing~\cite{grammarbasedFuzzing, whiteboxFuzzing} have been proposed (i.e., fuzzers also considering the source code of the program under test).
	However, the primary goal of fuzzing is not to generate test cases for regression testing 
	systematically traversing / revealing
	particular program modifications through 
	different possible paths and/or program versions as done in our approach.
	
	\paragraph{Plain Random.}
	A test-case generation technique that is similar to, yet simpler than, 
	is plain random test-case generation.~\cite{PRTest}.
	Test cases are randomly generated, and afterwards, 
	the achieved coverage is measured. 
	This approach is clearly more efficient than test-goal guided techniques.
	On the other hand, more complicated test goals are often not reached, 
	as the chance to generate valid input values reaching those goals is small 
	(e.g., to generate a test case for input value $x$ 
	which evaluates true for $x == 1$ is $\frac{1}{2^{32}}$ for a 32-bit system).
	Therefore, generating (multiple different) modification-revealing test cases is usually 
	extremely expensive and ineffective using purely random approaches. 
	
	\paragraph{Symbolic Execution.}
	Symbolic execution employs a symbolic reachability graph cope with
	the reachable state space of input programs during test-case generation.
	One prominent example is Klee~\cite{Klee}. 
	Based on symbolic execution, it might be also
	possible to generate multiple test cases covering the same goal through different paths as 
	done in our work. 
	However, we are not aware of any recent work going into this direction.
	
	\paragraph{Bounded Model Checking.}
	Another technique to scale test-case generation to larger
	programs is bounded model checking~\cite{ESBMC}.
	A bounded model checkers also computes the reachable state space of programs
	(either in an abstract or concrete representation), where 
	loops are only explored up to maximum number of iterations $k$ (bound).
	This enables the model checker to prove program properties with certainty only within that bound.
	Such a tool can be also used for test cases generation 
	similar to symbolic model checking (i.e., by encoding test goals as 
	reachbility problems, see below). 
	Again, we are not aware of any works using bounded model checking to generate multiple test cases for the same test goal
	or program modification, respectively.
	
	\paragraph{Symbolic Model Checking.}
	Another approach to handle larger
	input programs is symbolic model checking as applied, for instance, by the CPAchecker framework, which is also used by our approach~\cite{Beyer2004}.
	CoVeriTest is another recent test-case generator based on the CPAchecker framework~\cite{Coveritest}.
	Again, these and other tools do currently not support generation of 
	multiple test-cases per test goal or program modification as required in our approach.
	However, encoding the underlying problem as reachability query as done
	in our approach would also enable the usage of these other tools for
	regression-test generation.
	
	\subsubsection{Regression-Test-Case Generation.}
	We next discuss related work on generating test cases
	for systematically investigating semantic differences between similar programs.
	\emph{Differential testing}~\cite{McKeeman1998} is concerned with
	the following problem: Given two comparable programs 
	and an set of differential test cases, the systems can be checked for bugs by running the test cases. 
	If the outputs differ or the test loops indefinitely or crashes, 
	the test case is a candidate for a bug-revealing test.
	Thereupon, Evans and Savioa proposed an
	approach for detecting program changes by comparing test-execution results of two different program versions~\cite{Evans2007}.
	The tool \textsc{CSmith} combines differential testing with fuzzing (i.e., C programs) to find bugs in 
	C-compiler implementations~\cite{Yang2011}.
	The work being presumably most closely related to our methodology 
	is \textsc{DiffGen}~\cite{Taneja2008} for 
	generating test cases comparing two versions of a Java program.
	This is achieved by instrumenting
	programs with equality-assert-statements and generating test suites for statement coverage on (failed) assertions.
	This work differs from ours as it does not support multiple 
	test cases finding differences between program versions, and also does not 
	take multiple prior versions into account. 
	Additionally, they do not take TSR into account 
	to increase efficiency of regression testing.

	The goal of \emph{mutation testing} is to \emph{measure}
	effectiveness of test suites or test-generation techniques, respectively,
	by deriving from an original program a set of syntactically slightly changed \emph{mutants}
	(simulating faults)~\cite{Jia2011}.
	A test case \emph{detects} a mutant if its test-execution results
	for the original program differ from those for the mutant.
	Based on this principle, \cite{Fraser2010} pursue
	mutation-driven \emph{generation} of test cases for Java programs by using
	genetic algorithms to find test cases that detect mutants.
	In contrast, \cite{Harman2011SHOM} use a combination of symbolic execution and a search-based 
	heuristic to identify test cases that are likely to \emph{reveal} mutants.
	\cite{Souza2016} also propose a heuristic for generating 
	strong mutation-detecting test cases using hill climbing.
	As these works are mainly based on heuristics, the generated test cases
	do not guarantee to traverse/reveal program modifications.
	Additionally, those approaches 
	do not allow to configure the number of test cases or 
	the number of different programs as our parameters \textbf{NPR} and \textbf{NRT}, 
	nor do they consider TSR.

	Automatically \emph{generating} test cases for differentiating two
	program versions for regression testing has been initially proposed by \cite{Korel1998}.
	Their white-box testing tool automatically compares
	output values of two given program versions
	to derive input values leading to different outputs.
	Their approach is applicable to Pascal programs only and does
	not support multiple test cases and/or program versions as in our work.
	
	%
	\subsection{Regression Verification}
	The goal of regression verification is to analyze different versions of a system or program 
	to check whether the specification is still fulfilled after modifications. 
	To this end, intermediate verification results are re-used 
	between versions to increase efficiency~\cite{Hardin1996,Henzinger2003extreme,Strichman2008}.
	For instance, intermediate results (so-called \emph{abstraction precisions}) 
	of verification runs enable reuse for later version~\cite{Beyer2013regression}, 
	and regression verification may be applied to restrict the set of program inputs manually~\cite{Bohme2013} 
	or by syntactic checks~\cite{Godlin2013}.
	Furthermore, there is work on lifting principles of 
	regression verification to multi-threaded programs~\cite{Chaki2012} and 
	re-checking evolving software of automated production systems~\cite{Beckert2015}.
	Moreover, efficiency of regression verification may be improved, for instance, by applying 
	state-space partitioning-techniques~\cite{Backes2013} and by improved encodings of reuse-information~\cite{Felsing2014}.
	Other works in this area reuse final verification results 
	in case of a subsequent change in the program~\cite{Beyer2013,Visser2012,Sery2012,Yang2009,Lauterburg2008}.
	However, none of these approaches further utilizes the information collected for 
	regression analysis to derive modification-revealing test cases.

	Similarly, \emph{conditional model checking} aims at 
	reusing results of verification runs
	to perform collaborative verification~\cite{Beyer2012conditional,Christakis2012}.
	To this end, there are exchange formats for verification 
	witnesses for property violations~\cite{Beyer2015validation} as 
	well as for correctness proofs~\cite{Beyer2016correctness}.
	Furthermore, there is work on keeping track of 
	unverified parts of a program to apply test-case generation for these parts~\cite{Czech2015,Christakis2016}
	and reusing verification results for hybrid systems~\cite{Mitsch2014}.
	However, none of these approaches aim at finding differences between versions of the same 
	program as done in our approach.

	\section{Conclusion and Future Work}\label{sec7:conclusion}
	We presented a configurable regression-testing methodology for
	automating the selection of regression test-cases with a particular
	focus on revealing regression bugs in evolving programs. \textsc{RegreTS}
	currently supports regression testing of C programs at unit level.
	Our experimental results show that effectiveness and the efficiency
	highly depend on the selected regression test-case generation strategy,
	where the parameter $\textbf{RTC}$ (i.e., either generating modification-traversing or modification-revealing test cases) has the strongest
	impact on the effectiveness and efficiency.
	To conclude, our experimental results show that for obtaining the best efficiency in terms of CPU time and the best trade-off modification traversing test cases should be used without test-suite reduction.
	However, for the best effectiveness, modification revealing test cases should be used with multiple test cases per test goal and multiple previous revisions taken into account. Additionally, test-suite reduction obstructs effectiveness, therefore, for optimal effectiveness test-suite reduction should be disabled.
	
	As a future work, we plan to extend our approach. First, we
	plan to further improve our test-generation technique to support additional subject systems
	as well as in utilizing alternative test-case
	generation techniques within our methodology (e.g., symbolic execution) and to compare the outcome
	with our current results. Furthermore, we plan to investigate other
	kinds of regression errors and to identify suitable regression strategies
	for effectively revealing those bugs. Finally, we plan to adapt
	\textsc{RegreTS} to other testing scenarios including, for instance, other
	input languages besides C and other testing levels beyond unit
	testing which enables us to conduct experiments on a richer set of subject systems.
	
	\paragraph{Acknowledgments}
	This work was funded by the Hessian LOEWE initiative
	within the Software-Factory 4.0 project.
	
	\printbibliography

@InProceedings{Backes2013,
	author      =   {Backes, John and Person, Suzette and Rungta, Neha and Tkachuk, Oksana},
	booktitle   =   {SPIN '13},
	title       =   {{Regression Verification Using Impact Summaries}},
	year        =   {2013},
	volume      =   {7976},
	number      =   {},
	pages       =   {99--116},
	doi         =   {10.1007/978-3-642-39176-7_7},
	isbn        =   {978-3-642-39176-7},
	publisher   =   {Springer},
}

@InProceedings{Beckert2015,
	author      =   {Beckert, Bernhard and Ulbrich, Mattias and Vogel-Heuser, Birgit and Weigl, Alexander},
	booktitle   =   {ICFEM '15},
	title       =   {Regression Verification for Programmable Logic Controller Software},
	year        =   {2015},
	volume      =   {9407},
	number      =   {},
	pages       =   {234--251},
	doi         =   {10.1007/978-3-319-25423-4_15},
	isbn        =   {978-3-319-25423-4},
	publisher   =   {Springer International Publishing},
}

@inproceedings{Beyer2004,
	author      =   {Beyer, Dirk and Chlipala, Adam J. and Henzinger, Thomas A. and Jhala, Ranjit and Majumdar, Rupak},
	booktitle   =   {Proceedings of the 26th International Conference on Software Engineering},
	title       =   {Generating Tests from Counterexamples},
	year        =   {2004},
	volume      =   {},
	number      =   {},
	pages       =   {326--335},
	doi         =   {10.5555/998675.999437},
	isbn        =   {978-0769521633},
	publisher   =   {IEEE Computer Society},
}

@inproceedings{Beyer2012conditional,
	author      =   {Beyer, Dirk and Henzinger, Thomas A. and Keremoglu, M. Erkan and Wendler, Philipp},
	booktitle   =   {Proceedings of the ACM SIGSOFT 20th International Symposium on the Foundations of Software Engineering},
	title       =   {Conditional Model Checking: A Technique to Pass Information between Verifiers},
	year        =   {2012},
	volume      =   {},
	number      =   {},
	pages       =   {},
	doi         =   {10.1145/2393596.2393664},
	isbn        =   {9781450316149},
	publisher   =   {Association for Computing Machinery},
}

@InProceedings{Beyer2013,
	author      =   {Beyer, Dirk and Holzer, Andreas and Tautschnig, Michael and Veith, Helmut},
	booktitle   =   {ESOP '13},
	title       =   {Information Reuse for Multi-goal Reachability Analyses},
	year        =   {2013},
	volume      =   {7792},
	number      =   {},
	pages       =   {472--491},
	doi         =   {10.1007/978-3-642-37036-6_26},
	isbn        =   {978-3-642-37036-6},
	publisher   =   {Springer},
}

@inproceedings{Beyer2013regression,
	author      =   {Beyer, Dirk and L\"{o}we, Stefan and Novikov, Evgeny and Stahlbauer, Andreas and Wendler, Philipp},
	booktitle   =   {ESEC/FSE '13},
	title       =   {{Precision Reuse for Efficient Regression Verification}},
	year        =   {2013},
	volume      =   {},
	number      =   {},
	pages       =   {389--399},
	doi         =   {10.1145/2491411.2491429},
	isbn        =   {978-1-4503-2237-9},
	publisher   =   {ACM},
}

@inproceedings{Beyer2015validation,
	author      =   {Beyer, Dirk and Dangl, Matthias and Dietsch, Daniel and Heizmann, Matthias and Stahlbauer, Andreas},
	booktitle   =   {ESEC/FSE '15},
	title       =   {{Witness Validation and Stepwise Testification Across Software Verifiers}},
	year        =   {2015},
	volume      =   {},
	number      =   {},
	pages       =   {721--733},
	doi         =   {10.1145/2786805.2786867},
	isbn        =   {978-1-4503-3675-8},
	publisher   =   {ACM},
}

@inproceedings{Beyer2016correctness,
	author      =   {Beyer, Dirk and Dangl, Matthias and Dietsch, Daniel and Heizmann, Matthias},
	booktitle   =   {FSE '16},
	title       =   {{Correctness Witnesses: Exchanging Verification Results Between Verifiers}},
	year        =   {2016},
	volume      =   {},
	number      =   {},
	pages       =   {326--337},
	doi         =   {10.1145/2950290.2950351},
	isbn        =   {978-1-4503-4218-6},
	publisher   =   {ACM},
}

@inproceedings{Bohme2013,
	author      =   {B\"{o}hme, Marcel and Oliveira, Bruno C. d. S. and Roychoudhury, Abhik},
	booktitle   =   {ICSE '13},
	title       =   {{Partition-based Regression Verification}},
	year        =   {2013},
	volume      =   {},
	number      =   {},
	pages       =   {302--311},
	doi         =   {10.5555/2486788.2486829},
	isbn        =   {978-1-4673-3076-3},
	publisher   =   {IEEE Press},
}

@InProceedings{Chaki2012,
	author      =   {Chaki, Sagar and Gurfinkel, Arie and Strichman, Ofer},
	booktitle   =   {VMCAI '12},
	title       =   {{Regression Verification for Multi-threaded Programs}},
	year        =   {2012},
	volume      =   {7148},
	number      =   {},
	pages       =   {119--135},
	doi         =   {10.1007/978-3-642-27940-9_9},
	isbn        =   {978-3-642-27940-9},
	publisher   =   {Springer},
}

@InProceedings{Christakis2012,
	author      =   {Christakis, Maria and M{\"u}ller, Peter and W{\"u}stholz, Valentin},
	booktitle   =   {FM '12},
	title       =   {{Collaborative Verification and Testing with Explicit Assumptions}},
	year        =   {2012},
	volume      =   {7436},
	number      =   {},
	pages       =   {132--146},
	doi         =   {10.1007/978-3-642-32759-9_13},
	isbn        =   {978-3-642-32759-9},
	publisher   =   {Springer},
}

@inproceedings{Christakis2016,
	author      =   {Christakis, Maria and M\"{u}ller, Peter and W\"{u}stholz, Valentin},
	booktitle   =   {ICSE '16},
	title       =   {{Guiding Dynamic Symbolic Execution Toward Unverified Program Executions}},
	year        =   {2016},
	volume      =   {},
	number      =   {},
	pages       =   {144--155},
	doi         =   {10.1145/2884781.2884843},
	isbn        =   {978-1-4503-3900-1},
	publisher   =   {ACM},
}

@InProceedings{Czech2015,
	author      =   {Czech, Mike and Jakobs, Marie-Christine and Wehrheim, Heike},
	booktitle   =   {FASE '15},
	title       =   {{Just Test What You Cannot Verify!}},
	year        =   {2015},
	volume      =   {9033},
	number      =   {},
	pages       =   {100--114},
	doi         =   {10.1007/978-3-662-46675-9_7},
	isbn        =   {978-3-662-46675-9},
	publisher   =   {Springer},
}

@inproceedings{Engstrom2008,
	author      =   {Engstr\"{o}m, Emelie and Skoglund, Mats and Runeson, Per},
	booktitle   =   {Proceedings of the Second ACM-IEEE International Symposium on Empirical Software Engineering and Measurement},
	title       =   {{Empirical Evaluations of Regression Test Selection Techniques: A Systematic Review}},
	year        =   {2008},
	volume      =   {},
	number      =   {},
	pages       =   {22--31},
	doi         =   {10.1145/1414004.1414011},
	isbn        =   {978-1-59593-971-5},
	publisher   =   {ACM},
}

@inproceedings{Felsing2014,
	author      =   {Felsing, Dennis and Grebing, Sarah and Klebanov, Vladimir and R\"{u}mmer, Philipp and Ulbrich, Mattias},
	booktitle   =   {ASE '14},
	title       =   {{Automating Regression Verification}},
	year        =   {2014},
	volume      =   {},
	number      =   {},
	pages       =   {349--360},
	doi         =   {10.1145/2642937.2642987},
	isbn        =   {978-1-4503-3013-8},
	publisher   =   {ACM},
}

@inproceedings{Hardin1996,
	author      =   {Hardin, R. H. and Kurshan, R. P. and McMillan, K. L. and Reeds, J. A. and Sloane, N. J. A.},
	booktitle   =   {WODES '96},
	title       =   {Efficient regression verification},
	year        =   {1996},
	volume      =   {},
	number      =   {},
	pages       =   {157--150},
	doi         =   {},
	isbn        =   {},
	publisher   =   {IEE},
}

@INPROCEEDINGS{Lauterburg2008,
	author      =   {Lauterburg, Steven and Sobeih, Ahmed and Marinov, Darko and Viswanathan, Mahesh},
	booktitle   =   {ICSE '08},
	title       =   {{Incremental State-Space Exploration for Programs with Dynamically Allocated Data}},
	year        =   {2008},
	volume      =   {},
	number      =   {},
	pages       =   {291--300},
	doi         =   {10.1145/1368088.1368128},
	isbn        =   {},
	publisher   =   {ACM},
}

@INPROCEEDINGS{Sery2012,
	author      =   {Sery, Ondrej and Fedyukovich, Grigory and Sharygina, Natasha},
	booktitle   =   {FMCAD '12},
	title       =   {{Incremental Upgrade Checking by Means of Interpolation-based Function Summaries}},
	year        =   {2012},
	volume      =   {},
	number      =   {},
	pages       =   {114--121},
	doi         =   {},
	isbn        =   {},
	publisher   =   {IEEE},
}

@inproceedings{Visser2012,
	author      =   {Visser, Willem and Geldenhuys, Jaco and Dwyer, Matthew B.},
	booktitle   =   {FSE '12},
	title       =   {{Green: Reducing, Reusing and Recycling Constraints in Program Analysis}},
	year        =   {2012},
	volume      =   {},
	number      =   {},
	pages       =   {58:1--58:11},
	doi         =   {10.1145/2393596.2393665},
	isbn        =   {978-1-4503-1614-9},
	publisher   =   {ACM},
}

@INPROCEEDINGS{Yang2009,
	author      =   {Yang, Guowei and Dwyer, Matthew B. and Rothermel, Gregg},
	booktitle   =   {ICSM '19},
	title       =   {{Regression Model Checking}},
	year        =   {2009},
	volume      =   {},
	number      =   {},
	pages       =   {115--124},
	doi         =   {10.1109/ICSM.2009.5306334},
	isbn        =   {},
	publisher   =   {IEEE},
}

@inproceedings{Yang2011,
	author      =   {Yang, Xuejun and Chen, Yang and Eide, Eric and Regehr, John},
	booktitle   =   {Proceedings of the 32Nd ACM SIGPLAN Conference on Programming Language Design and Implementation},
	title       =   {Finding and Understanding Bugs in C Compilers},
	year        =   {2011},
	volume      =   {},
	number      =   {},
	pages       =   {283--294},
	doi         =   {10.1145/1993498.1993532},
	isbn        =   {978-1-4503-0663-8},
	publisher   =   {ACM},
}

@inproceedings{Evans2007,
	author      =   {Evans, Robert B. and Savoia, Alberto},
	booktitle   =   {The 6th Joint Meeting on European Software Engineering Conference and the ACM SIGSOFT Symposium on the Foundations of Software Engineering: Companion Papers},
	title       =   {Differential Testing: A New Approach to Change Detection},
	year        =   {2007},
	volume      =   {},
	number      =   {},
	pages       =   {549--552},
	doi         =   {10.1145/1295014.1295038},
	isbn        =   {978-1-59593-812-1},
	publisher   =   {ACM},
}

@InProceedings{Taneja2008,
	author      =   {Kunal {Taneja} and Tao {Xie}},
	booktitle   =   {2008 23rd IEEE/ACM International Conference on Automated Software Engineering},
	title       =   {DiffGen: Automated Regression Unit-Test Generation},
	year        =   {2008},
	volume      =   {},
	number      =   {},
	pages       =   {407--410},
	doi         =   {10.1109/ASE.2008.60},
	isbn        =   {},
	publisher   =   {IEEE},
}

@inproceedings{Fraser2010,
	author      =   {Fraser, Gordon and Zeller, Andreas},
	booktitle   =   {Proceedings of the 19th International Symposium on Software Testing and Analysis},
	title       =   {Mutation-driven Generation of Unit Tests and Oracles},
	year        =   {2010},
	volume      =   {},
	number      =   {},
	pages       =   {147--158},
	doi         =   {10.1145/1831708.1831728},
	isbn        =   {978-1-60558-823-0},
	publisher   =   {ACM},
}

@inproceedings{Harman2011SHOM,
	author      =   {Harman, Mark and Jia, Yue and Langdon, William B.},
	booktitle   =   {Proceedings of the 19th ACM SIGSOFT Symposium and the 13th European Conference on Foundations of Software Engineering},
	title       =   {Strong Higher Order Mutation-based Test Data Generation},
	year        =   {2011},
	volume      =   {},
	number      =   {},
	pages       =   {212--222},
	doi         =   {10.1145/2025113.2025144},
	isbn        =   {978-1-4503-0443-6},
	publisher   =   {ACM},
}

@inproceedings{music,
	author      =   {Duy Loc Phan and Yunho Kim and Moonzoo Kim},
	booktitle   =   {2018 {IEEE} International Conference on Software Testing, Verification and Validation Workshops},
	title       =   {{MUSIC:} Mutation Analysis Tool with High Configurability and Extensibility},
	year        =   {2018},
	volume      =   {},
	number      =   {},
	pages       =   {40--46},
	doi         =   {10.1109/ICSTW.2018.00026},
	isbn        =   {},
	publisher   =   {IEEE},
}

@inproceedings{Korel1998,
	author      =   {Korel, Bogdan and Al-Yami, Ali M.},
	booktitle   =   {Proceedings of the 1998 ACM SIGSOFT International Symposium on Software Testing and Analysis},
	title       =   {Automated Regression Test Generation},
	year        =   {1998},
	volume      =   {},
	number      =   {},
	pages       =   {143--152},
	doi         =   {10.1145/271771.271803},
	isbn        =   {0-89791-971-8},
	publisher   =   {ACM},
}

@inproceedings{Souza2016,
	author      =   {Souza, Francisco Carlos M. and Papadakis, Mike and Le Traon, Yves and Delamaro, M\'{a}rcio E.},
	booktitle   =   {Proceedings of the 9th International Workshop on Search-Based Software Testing},
	title       =   {Strong Mutation-based Test Data Generation Using Hill Climbing},
	year        =   {2016},
	volume      =   {},
	number      =   {},
	pages       =   {45--54},
	doi         =   {10.1145/2897010.2897012},
	isbn        =   {978-1-4503-4166-0},
	publisher   =   {ACM},
}

@InProceedings{Horgan1992,
	author      =   {Joseph R. {Horgan} and Saul {London}},
	booktitle   =   {[1992] Proceedings of the Second Symposium on Assessment of Quality Software Development Tools},
	title       =   {A data flow coverage testing tool for C},
	year        =   {1992},
	volume      =   {},
	number      =   {},
	pages       =   {2--10},
	doi         =   {10.1109/AQSDT.1992.205829},
	isbn        =   {},
	publisher   =   {IEEE},
}

@INPROCEEDINGS{Offutt1995,
	author      =   {A. Jefferson Offutt and Jie Pan and Jeffrey M. Voas},
	booktitle   =   {Twelfth International Conference on Testing Computer Software},
	title       =   {Procedures for reducing the size of coverage-based test sets},
	year        =   {1995},
	volume      =   {},
	number      =   {},
	pages       =   {111--123},
	doi         =   {},
	isbn        =   {},
	publisher   =   {},
}

@InProceedings{Hartmann1989,
	author      =   {J. {Hartmann} and David J. {Robson}},
	booktitle   =   {Proceedings. Conference on Software Maintenance - 1989},
	title       =   {Revalidation during the software maintenance phase},
	year        =   {1989},
	volume      =   {},
	number      =   {},
	pages       =   {70--80},
	doi         =   {10.1109/ICSM.1989.65195},
	isbn        =   {},
	publisher   =   {IEEE},
}

@InProceedings{Hartmann1990,
	author      =   {J. {Hartmann} and David J. {Robson}},
	booktitle   =   {Twenty-Third Annual Hawaii International Conference on System Sciences},
	title       =   {RETEST-development of a selective revalidation prototype environment for use in software maintenance},
	year        =   {1990},
	volume      =   {2},
	number      =   {},
	pages       =   {92-101 vol.2},
	doi         =   {10.1109/HICSS.1990.205179},
	isbn        =   {},
	publisher   =   {IEEE},
}

@InProceedings{Gupta1992,
	author      =   {Rajiv {Gupta} and Mary Jean {Harrold} and Mary Lou {Soffa}},
	booktitle   =   {Proceedings Conference on Software Maintenance 1992},
	title       =   {An approach to regression testing using slicing},
	year        =   {1992},
	volume      =   {},
	number      =   {},
	pages       =   {299--308},
	doi         =   {10.1109/ICSM.1992.242531},
	isbn        =   {},
	publisher   =   {IEEE Computer Society},
}

@InProceedings{Harrold1988,
	author      =   {Mary Jean {Harrold} and M. L. {Souffa}},
	booktitle   =   {Proceedings. Conference on Software Maintenance, 1988.},
	title       =   {An incremental approach to unit testing during maintenance},
	year        =   {1988},
	volume      =   {},
	number      =   {},
	pages       =   {362--367},
	doi         =   {10.1109/ICSM.1988.10188},
	isbn        =   {},
	publisher   =   {IEEE},
}

@inproceedings{Harrold1989Interprocedual,
	author      =   {Harrold, Mary Jean and Soffa, Mary Lou},
	booktitle   =   {Proceedings of the ACM SIGSOFT '89 Third Symposium on Software Testing, Analysis, and Verification},
	title       =   {Interprocedual Data Flow Testing},
	year        =   {1989},
	volume      =   {},
	number      =   {},
	pages       =   {158--167},
	doi         =   {10.1145/75308.75327},
	isbn        =   {0-89791-342-6},
	publisher   =   {ACM},
}

@InProceedings{Taha1989,
	author      =   {Abu-Bakr {Taha} and Stephen M. {Thebaut} and Sying-Syang {Liu}},
	booktitle   =   {[1989] Proceedings of the Thirteenth Annual International Computer Software Applications Conference},
	title       =   {An approach to software fault localization and revalidation based on incremental data flow analysis},
	year        =   {1989},
	volume      =   {},
	number      =   {},
	pages       =   {527--534},
	doi         =   {10.1109/CMPSAC.1989.65142},
	isbn        =   {},
	publisher   =   {IEEE},
}

@InProceedings{Wong1995,
	author      =   {Weichen E. {Wong} and Joseph R. {Horgan} and Saul {London} and Aditya P. {Mathur}},
	booktitle   =   {1995 17th International Conference on Software Engineering},
	title       =   {Effect of Test Set Minimization on Fault Detection Effectiveness},
	year        =   {1995},
	volume      =   {},
	number      =   {},
	pages       =   {41--41},
	doi         =   {10.1145/225014.225018},
	isbn        =   {},
	publisher   =   {ACM},
}

@InProceedings{Rothermel1999,
	author      =   {Gregg {Rothermel} and Roland H. {Untch} and {Chengyun Chu} and Mary Jean {Harrold}},
	booktitle   =   {Proceedings IEEE International Conference on Software Maintenance - 1999 (ICSM'99). 'Software Maintenance for Business Change' (Cat. No.99CB36360)},
	title       =   {Test case prioritization: an empirical study},
	year        =   {1999},
	volume      =   {},
	number      =   {},
	pages       =   {179--188},
	doi         =   {10.1109/ICSM.1999.792604},
	isbn        =   {},
	publisher   =   {IEEE},
}

@InProceedings{Elbaum2001,
	author      =   {Sebastian {Elbaum} and David {Gable} and Gregg {Rothermel}},
	booktitle   =   {Proceedings Seventh International Software Metrics Symposium},
	title       =   {Understanding and measuring the sources of variation in the prioritization of regression test suites},
	year        =   {2001},
	volume      =   {},
	number      =   {},
	pages       =   {169--179},
	doi         =   {10.1109/METRIC.2001.915525},
	isbn        =   {},
	publisher   =   {IEEE Computer Society},
}

@InProceedings{Elbaum2001Costs,
	author      =   {Sebastian {Elbaum} and Alexey {Malishevsky} and Gregg {Rothermel}},
	booktitle   =   {Proceedings of the 23rd International Conference on Software Engineering. ICSE 2001},
	title       =   {Incorporating varying test costs and fault severities into test case prioritization},
	year        =   {2001},
	volume      =   {},
	number      =   {},
	pages       =   {329--338},
	doi         =   {10.1109/ICSE.2001.919106},
	isbn        =   {},
	publisher   =   {IEEE Computer Society},
}

@InProceedings{Malishevsky2002,
	author      =   {Alexey {Malishevsky} and Gregg {Rothermel} and Sebastian {Elbaum}},
	booktitle   =   {International Conference on Software Maintenance, 2002. Proceedings.},
	title       =   {Modeling the cost-benefits tradeoffs for regression testing techniques},
	year        =   {2002},
	volume      =   {},
	number      =   {},
	pages       =   {204--213},
	doi         =   {10.1109/ICSM.2002.1167767},
	isbn        =   {},
	publisher   =   {IEEE},
}

@InProceedings{Rothermel2002,
	author      =   {Gregg {Rothermel} and Sebastian {Elbaum} and Alexey {Malishevsky} and Praveen {Kallakuri} and Brian {Davia}},
	booktitle   =   {Proceedings of the 24th International Conference on Software Engineering. ICSE 2002},
	title       =   {The impact of test suite granularity on the cost-effectiveness of regression testing},
	year        =   {2002},
	volume      =   {},
	number      =   {},
	pages       =   {130--140},
	doi         =   {10.1145/581356.581358},
	isbn        =   {},
	publisher   =   {ACM},
}

@InProceedings{Wang2016,
	author      =   {Xiaolin {Wang} and Hongwei {Zeng}},
	booktitle   =   {2016 IEEE/ACM International Workshop on Continuous Software Evolution and Delivery (CSED)},
	title       =   {History-Based Dynamic Test Case Prioritization for Requirement Properties in Regression Testing},
	year        =   {2016},
	volume      =   {},
	number      =   {},
	pages       =   {41--47},
	doi         =   {10.1109/CSED.2016.016},
	isbn        =   {},
	publisher   =   {IEEE},
}

@inproceedings{Wang2018,
	author      =   {Wang, Kaiyuan and Zhu, Chenguang and Celik, Ahmet and Kim, Jongwook and Batory, Don and Gligoric, Milos},
	booktitle   =   {Proceedings of the 40th International Conference on Software Engineering},
	title       =   {Towards Refactoring-aware Regression Test Selection},
	year        =   {2018},
	volume      =   {},
	number      =   {},
	pages       =   {233--244},
	doi         =   {10.1145/3180155.3180254},
	isbn        =   {978-1-4503-5638-1},
	publisher   =   {ACM},
}

@inproceedings{Marijan2018,
	author = {Marijan, Dusica and Liaaen, Marius},
	booktitle = {Proceedings of the 40th International Conference on Software Engineering: Software Engineering in Practice},
	title = {Practical Selective Regression Testing with Effective Redundancy in Interleaved Tests},
	year = {2018},
	volume = {},
	numer = {},
	pages = {153–162},
	doi = {10.1145/3183519.3183532},
	isbn = {9781450356596},
	publisher = {Association for Computing Machinery},
}

@inproceedings{Choudhary2018,
	author      =   {Choudhary, Ankur and Agrawal, Arun Prakash and Kaur, Arvinder},
	booktitle   =   {Proceedings of the 11th International Workshop on Search-Based Software Testing},
	title       =   {An Effective Approach for Regression Test Case Selection Using Pareto Based Multi-objective Harmony Search},
	year        =   {2018},
	volume      =   {},
	number      =   {},
	pages       =   {13--20},
	doi         =   {10.1145/3194718.3194722},
	isbn        =   {978-1-4503-5741-8},
	publisher   =   {ACM},
}

@InProceedings{Andrews2005,
	author      =   {James H. {Andrews} and Lionel C. {Briand} and Yvan {Labiche}},
	booktitle   =   {Proceedings. 27th International Conference on Software Engineering, 2005. ICSE 2005.},
	title       =   {Is mutation an appropriate tool for testing experiments? [software testing]},
	year        =   {2005},
	volume      =   {},
	number      =   {},
	pages       =   {402--411},
	doi         =   {10.1109/ICSE.2005.1553583},
	isbn        =   {},
	publisher   =   {IEEE},
}

@InProceedings{corebench,
	author      =   {B\"{o}hme, Marcel and Roychoudhury, Abhik},
	booktitle   =   {Proceedings of the 23rd ACM/SIGSOFT International Symposium on Software Testing and Analysis},
	title       =   {CoREBench: Studying Complexity of Regression Errors},
	year        =   {2014},
	volume      =   {},
	number      =   {},
	pages       =   {105--115},
	doi         =   {},
	isbn        =   {},
	publisher   =   {ACM},
}

@InProceedings{Chen_2017,
	author      =   {Junjie Chen and Yanwei Bai and Dan Hao and Lingming Zhang and Lu Zhang and Bing Xie},
	booktitle   =   {2017 {IEEE} International Conference on Software Testing, Verification and Validation ({ICST})},
	title       =   {How Do Assertions Impact Coverage-Based Test-Suite Reduction?},
	year        =   {2017},
	volume      =   {},
	number      =   {},
	pages       =   {},
	doi         =   {10.1109/icst.2017.45},
	isbn        =   {},
	publisher   =   {{IEEE}},
}

@InProceedings{Shi_2018,
	author      =   {Shi, August and Gyori, Alex and Mahmood, Suleman and Zhao, Peiyuan and Marinov, Darko},
	booktitle   =   {Proceedings of the 27th ACM SIGSOFT International Symposium on Software Testing and Analysis},
	title       =   {Evaluating Test-Suite Reduction in Real Software Evolution},
	year        =   {2018},
	volume      =   {},
	number      =   {},
	pages       =   {84--94},
	doi         =   {10.1145/3213846.3213875},
	isbn        =   {9781450356992},
	publisher   =   {Association for Computing Machinery},
}

@InProceedings{Shi_2015,
	author      =   {Shi, August and Yung, Tifany and Gyori, Alex and Marinov, Darko},
	booktitle   =   {Proceedings of the 2015 10th Joint Meeting on Foundations of Software Engineering},
	title       =   {Comparing and Combining Test-Suite Reduction and Regression Test Selection},
	year        =   {2015},
	volume      =   {},
	number      =   {},
	pages       =   {237--247},
	doi         =   {10.1145/2786805.2786878},
	isbn        =   {9781450336758},
	publisher   =   {Association for Computing Machinery},
}

@InProceedings{Khalilian2012,
	author      =   {Khalilian, Alireza and Parsa, Saeed},
	booktitle   =   {Advances in Software Engineering Techniques},
	title       =   {Bi-criteria Test Suite Reduction by Cluster Analysis of Execution Profiles},
	year        =   {2012},
	volume      =   {},
	number      =   {},
	pages       =   {243--256},
	doi         =   {},
	isbn        =   {978-3-642-28038-2},
	publisher   =   {Springer Berlin Heidelberg},
}

@InProceedings{Inozemtseva2014,
	author      =   {Inozemtseva, Laura and Holmes, Reid},
	booktitle   =   {Proceedings of the 36th International Conference on Software Engineering},
	title       =   {Coverage is Not Strongly Correlated with Test Suite Effectiveness},
	year        =   {2014},
	volume      =   {},
	number      =   {},
	pages       =   {435--445},
	doi         =   {10.1145/2568225.2568271},
	isbn        =   {9781450327565},
	publisher   =   {Association for Computing Machinery},
}

@InProceedings{Kim2002,
	author      =   {Jung-Min Kim and Porter, A.},
	booktitle   =   {Proceedings of the 24th International Conference on Software Engineering. ICSE 2002},
	title       =   {A history-based test prioritization technique for regression testing in resource constrained environments},
	year        =   {2002},
	volume      =   {},
	number      =   {},
	pages       =   {119--129},
	doi         =   {10.1109/ICSE.2002.1007961},
	isbn        =   {158113472X},
	publisher   =   {Association for Computing Machinery},
}

@InProceedings{Cruciani2019,
	author      =   {Cruciani, Emilio and Miranda, Breno and Verdecchia, Roberto and Bertolino, Antonia},
	booktitle   =   {2019 IEEE/ACM 41st International Conference on Software Engineering (ICSE)},
	title       =   {Scalable Approaches for Test Suite Reduction},
	year        =   {2019},
	volume      =   {},
	number      =   {},
	pages       =   {419--429},
	doi         =   {10.1109/ICSE.2019.00055},
	isbn        =   {},
	publisher   =   {IEEE Press},
}

@inproceedings{TestCov,
	author      =   {Beyer, Dirk and Lemberger, Thomas},
	booktitle   =   {Proceedings of the 34th IEEE/ACM International Conference on Automated Software Engineering},
	title       =   {TestCov: Robust Test-Suite Execution and Coverage Measurement},
	year        =   {2019},
	volume      =   {},
	number      =   {},
	pages       =   {1074–1077},
	doi         =   {10.1109/ASE.2019.00105},
	isbn        =   {9781728125084},
	publisher   =   {IEEE Press},
}

@InProceedings{evolutionaryFuzzing,
	author      =   {Sanjay Rawat and Vivek Jain and Ashish Kumar and Lucian Cojocar and Cristiano Giuffrida and Herbert Bos},
	booktitle   =   {Proceedings 2017 Network and Distributed System Security Symposium},
	title       =   {{VUzzer}: Application-aware Evolutionary Fuzzing},
	year        =   {2017},
	volume      =   {},
	number      =   {},
	pages       =   {},
	doi         =   {10.14722/ndss.2017.23404},
	isbn        =   {},
	publisher   =   {Internet Society},
}

@inproceedings{grammarbasedFuzzing,
	author      =   {Godefroid, Patrice and Kiezun, Adam and Levin, Michael Y.},
	booktitle   =   {Proceedings of the 29th ACM SIGPLAN Conference on Programming Language Design and Implementation},
	title       =   {Grammar-Based Whitebox Fuzzing},
	year        =   {2008},
	volume      =   {},
	number      =   {},
	pages       =   {206–215},
	doi         =   {10.1145/1375581.1375607},
	isbn        =   {9781595938602},
	publisher   =   {Association for Computing Machinery},
}

@InProceedings{whiteboxFuzzing,
	author      =   {Godefroid, Patrice and Levin, Michael Y. and Molnar, David},
	booktitle   =   {Network Distributed Security Symposium (NDSS)},
	title       =   {Automated Whitebox Fuzz Testing},
	year        =   {2008},
}

@inproceedings {Klee,
	author      =   {Cristian Cadar and Daniel Dunbar and Dawson Engler},
	booktitle   =   {8th {USENIX} Symposium on Operating Systems Design and Implementation ({OSDI} 08)},
	title       =   {{KLEE}: Unassisted and Automatic Generation of High-Coverage Tests for Complex Systems Programs},
	year        =   {2008},
	volume      =   {8},
	number      =   {},
	pages       =   {209–224},
	doi         =   {},
	isbn        =   {},
	publisher   =   {{USENIX} Association},
}

@InProceedings{ESBMC,
	author      =   {Gadelha, Mikhail R. and Menezes, Rafael and Monteiro, Felipe R. and Cordeiro, Lucas C.	and Nicole, Denis},
	booktitle   =   {Fundamental Approaches to Software Engineering},
	title       =   {{ESBMC}: Scalable and Precise Test Generation based on the Floating-Point Theory},
	year        =   {2020},
	volume      =   {},
	number      =   {},
	pages       =   {525--529},
	doi         =   {10.1007/978-3-030-45234-6_27},
	isbn        =   {978-3-030-45234-6},
	publisher   =   {Springer International Publishing},
}

@InProceedings{Coveritest,
	author      =   {Jakobs, Marie-Christine},
	booktitle   =   {Fundamental Approaches to Software Engineering},
	title       =   {{CoVeriTest} with Dynamic Partitioning of the Iteration Time Limit (Competition Contribution)},
	year        =   {2020},
	volume      =   {12076},
	number      =   {},
	pages       =   {540--544},
	doi         =   {10.1007/978-3-030-45234-6\_30},
	isbn        =   {978-3-030-45234-6},
	publisher   =   {Springer International Publishing},
}

@InProceedings{Testcomp21,
	author      =   {Beyer, Dirk},
	booktitle   =   {Fundamental Approaches to Software Engineering},
	title       =   {Status Report on Software Testing: Test-Comp 2021},
	year        =   {2021},
	volume      =   {},
	number      =   {},
	pages       =   {341--357},
	doi         =   {10.1007/978-3-030-71500-7_17},
	isbn        =   {978-3-030-71500-7},
	publisher   =   {Springer International Publishing},
}

@Book{Ammann2016,
	year        =   {2016},
	author      =   {Ammann, Paul and Offutt, Jeff},
	isbn        =   {978-1-107-17201-2},
	publisher   =   {Cambridge University Press},
	title       =   {{Introduction to Software Testing}},
}

@Inbook{Henzinger2003extreme,
	volume      =   {2772},
	pages       =   {332--358},
	year        =   {2003},
	author      =   {Henzinger, Thomas A. and Jhala, Ranjit and Majumdar, Rupak and Sanvido, Marco A. A.},
	series      =   {LNCS},
	isbn        =   {978-3-540-39910-0},
	publisher   =   {Springer},
	title       =   {{Extreme Model Checking}},
	booktitle   =   {Verification: Theory and Practice},
	doi         =   {10.1007/978-3-540-39910-0_16},
}

@Inbook{Strichman2008,
	volume      =   {4171},
	pages       =   {496--501},
	year        =   {2008},
	author      =   {Strichman, Ofer and Godlin, Benny},
	series      =   {LNCS},
	isbn        =   {978-3-540-69149-5},
	publisher   =   {Springer},
	title       =   {{Regression Verification - A Practical Way to Verify Programs}},
	booktitle   =   {VSTTE '05},
	doi         =   {10.1007/978-3-540-69149-5_54},
}

@article{Godlin2013,
	author      =   {Godlin, Benny and Strichman, Ofer},
	journal     =   {Software Testing, Verification and Reliability},
	title       =   {Regression verification: proving the equivalence of similar programs},
	year        =   {2013},
	volume      =   {23},
	number      =   {3},
	pages       =   {241--258},
	doi         =   {10.1002/stvr.1472},
	issn        =   {},
	publisher   =   {},
}

@Article{Mitsch2014,
	author      =   {Mitsch, Stefan and Passmore, Grant Olney and Platzer, Andr{\'e}},
	journal     =   {Mathematics in Computer Science},
	title       =   {Collaborative Verification-Driven Engineering of Hybrid Systems},
	year        =   {2014},
	volume      =   {8},
	number      =   {1},
	pages       =   {71--97},
	doi         =   {10.1007/s11786-014-0176-y},
	issn        =   {1661-8289},
	publisher   =   {},
}

@article{Yoo2012,
	author      =   {Yoo, Shin and Harman, Mark},
	journal     =   {Softw. Test. Verif. Reliab.},
	title       =   {Regression Testing Minimization, Selection and Prioritization: A Survey},
	year        =   {2012},
	volume      =   {22},
	number      =   {2},
	pages       =   {67--120},
	doi         =   {10.1002/stv.430},
	issn        =   {0960-0833},
	publisher   =   {John Wiley and Sons Ltd.},
}

@article{Chen1996,
	author      =   {Chen, Tsong Yueh and Lau, Man Fai},
	journal     =   {Inf. Process. Lett.},
	title       =   {Dividing Strategies for the Optimization of a Test Suite},
	year        =   {1996},
	volume      =   {60},
	number      =   {3},
	pages       =   {135--141},
	doi         =   {10.1016/S0020-0190(96)00135-4},
	issn        =   {0020-0190},
	publisher   =   {Elsevier North-Holland, Inc.},
}

@Article{McKeeman1998,
	author      =   {William M. McKeeman},
	journal     =   {Digital Technical Journal},
	title       =   {Differential Testing for Software},
	year        =   {1998},
	volume      =   {10},
	number      =   {},
	pages       =   {100--107},
	doi         =   {},
	issn        =   {},
	publisher   =   {},
}

@Article{Anand2013,
	author      =   {Saswat Anand and Edmund K. Burke and Tsong Yueh Chen and John Clark and Myra B. Cohen and Wolfgang Grieskamp and Mark Harman and Mary Jean Harrold and Phil McMinn and Antonia Bertolino and J. {Jenny Li} and Hong Zhu},
	journal     =   {Journal of Systems and Software},
	title       =   {An orchestrated survey of methodologies for automated software test case generation},
	year        =   {2013},
	volume      =   {86},
	number      =   {8},
	pages       =   {1978--2001},
	doi         =   {https://doi.org/10.1016/j.jss.2013.02.061},
	issn        =   {0164-1212},
	publisher   =   {},
}

@Article{Pecorelli2021,
	author      =   {Fabiano Pecorelli and Fabio Palomba and Andrea De Lucia},
	journal     =   {Empirical Software Engineering},
	title       =   {The Relation of Test-Related Factors to Software Quality: A Case Study on Apache Systems},
	year        =   {2021},
	volume      =   {26},
	number      =   {2},
	pages       =   {},
	doi         =   {10.1007/s10664-020-09891-y},
	issn        =   {},
	publisher   =   {Springer Science and Business Media {LLC}},
}

@article{sir,
	author      =   {Hyunsook Do and Sebastian G. Elbaum and Gregg Rothermel},
	journal     =   {Empirical Software Engineering: An International Journal},
	title       =   {Supporting Controlled Experimentation with Testing Techniques},
	year        =   {2005},
	volume      =   {10},
	number      =   {4},
	pages       =   {405--435},
	doi         =   {},
	issn        =   {},
	publisher   =   {},
}

@misc{svbenchmarks,
	year        =   {2019},
	author      =   {Beyer, Dirk},
	publisher   =   {Zenodo},
	title       =   {SV-Benchmarks: Benchmark Set of 8th Intl. Competition on Software Verification (SV-COMP 2019)},
	doi         =   {10.5281/zenodo.2598728},
}

@Article{Jia2011,
	author      =   {Yue Jia and Mark Harman},
	journal     =   {IEEE Transactions on Software Engineering},
	title       =   {{An Analysis and Survey of the Development of Mutation Testing}},
	year        =   {2011},
	volume      =   {37},
	number      =   {5},
	pages       =   {649--678},
	doi         =   {10.1109/TSE.2010.62},
	issn        =   {0098-5589},
	publisher   =   {},
}

@Article{Rothermel2001,
	author      =   {Gregg {Rothermel} and Roland H. {Untch} and {Chengyun Chu} and Mary J. {Harrold}},
	journal     =   {IEEE Transactions on Software Engineering},
	title       =   {Prioritizing test cases for regression testing},
	year        =   {2001},
	volume      =   {27},
	number      =   {10},
	pages       =   {929--948},
	doi         =   {10.1109/32.962562},
	issn        =   {0098-5589},
	publisher   =   {},
}

@Article{Hartmann1990Revalidation,
	author      =   {J. {Hartmann} and David J. {Robson}},
	journal     =   {IEEE Software},
	title       =   {Techniques for selective revalidation},
	year        =   {1990},
	volume      =   {7},
	number      =   {1},
	pages       =   {31--36},
	doi         =   {10.1109/52.43047},
	issn        =   {0740-7459},
	publisher   =   {},
}

@article{Harrold1993,
	author      =   {Harrold, Mary Jean and Gupta, Rajiv and Soffa, Mary Lou},
	journal     =   {ACM Trans. Softw. Eng. Methodol.},
	title       =   {A Methodology for Controlling the Size of a Test Suite},
	year        =   {1993},
	volume      =   {2},
	number      =   {3},
	pages       =   {270--285},
	doi         =   {10.1145/152388.152391},
	issn        =   {1049-331X},
	publisher   =   {ACM},
}

@Article{FuzzingOverview,
	author      =   {Jun Li and Bodong Zhao and Chao Zhang},
	journal     =   {Cybersecurity},
	title       =   {Fuzzing: a survey},
	year        =   {2018},
	volume      =   {1},
	number      =   {1},
	pages       =   {},
	doi         =   {10.1186/s42400-018-0002-y},
	issn        =   {},
	publisher   =   {Springer Science and Business Media {LLC}},
}

@Article{PRTest,
	author      =   {Thomas Lemberger},
	journal     =   {International Journal on Software Tools for Technology Transfer},
	title       =   {Plain random test generation with {PRTest}},
	year        =   {2020},
	volume      =   {},
	number      =   {},
	pages       =   {},
	doi         =   {10.1007/s10009-020-00568-x},
	issn        =   {},
	publisher   =   {Springer Science and Business Media {LLC}},
}

@TechReport{aflFuzzer,
	author      =   {Michał Zalewski},
	title       =   {Technical "whitepaper" for afl-fuzz},
	year        =   {2006},
	url         =   {https://lcamtuf.coredump.cx/afl/technical_details.txt},
}

@inproceedings{DegradedILP2008,
	author = {Chen, Zhenyu and Zhang, Xiaofang and Xu, Baowen},
	year = {2008},
	month = {01},
	pages = {494-499},
	title = {A Degraded ILP Approach for Test Suite Reduction.},
	journal = {20th International Conference on Software Engineering and Knowledge Engineering, SEKE 2008},
	doi = {},
}
	
\end{document}